\documentclass[aps,twocolumn,showpacs,preprintnumbers,amsmath,amssymb,nofootinbib]{revtex4}
\usepackage{graphicx}
\usepackage{dcolumn}
\usepackage{bm}

\def\simge{\mathrel{
     \rlap{\raise 0.511ex \hbox{$>$}}{\lower 0.511ex \hbox{$\sim$}}}}
\def\simle{\mathrel{
     \rlap{\raise 0.511ex \hbox{$<$}}{\lower 0.511ex \hbox{$\sim$}}}}
\def\be{\begin{equation}}
\def\ee{\end{equation}}
\def\bea{\begin{eqnarray}}
\def\eea{\end{eqnarray}}

\newcommand{\calo}{{\cal O}}
\newcommand{\boldl}{{\bf L}}
\newcommand{\boldu}{{\bf U}}
\newcommand{\boldx}{{\bf x}}
\newcommand{\tr}{{\rm tr}\,}
\newcommand{\Tr}{{\rm Tr}\,}

\preprint{BNL-101281-2013-JA, RBRC 1030}

\begin{document}
\title{Effective potential for SU(2) Polyakov loops and Wilson loop eigenvalues}

\author{Dominik Smith,$^{a}$ Adrian Dumitru,$^{b,c}$ 
Robert Pisarski,$^{c,d}$ Lorenz von Smekal$^{a,e}$}

\affiliation{
$^a$Theoriezentrum, Institut f\"ur Kernphysik, TU Darmstadt,
   64289 Darmstadt, Germany\\ 
$^b$Department of Natural Sciences, Baruch College, New York, NY 10010, USA\\
$^c$RIKEN/BNL Research Center, Brookhaven National Laboratory, Upton, NY 11973, USA\\
$^d$Department of Physics, Brookhaven National Laboratory, Upton, NY
  11973, USA\\
$^e$Institut f\"ur Theoretische Physik, Justus-Liebig-Universit\"at,
   35392 Giessen, Germany 
}

\date{\today}
\begin{abstract}
We simulate SU(2) gauge theory at temperatures ranging from slightly
below $T_c$ to roughly $2T_c$ for two different values of the gauge
coupling. Using a histogram method, we extract the effective potential
for the Polyakov loop and for the phases of the eigenvalues of the
thermal Wilson loop, in both the fundamental and adjoint
representations.  We show that the classical potential of the
fundamental loop can be parametrized within a simple model which
includes a Vandermonde potential and terms linear and quadratic in the
Polyakov loop. We discuss how parametrizations for the other cases can
be obtained from this model.
\end{abstract}
\maketitle

\section{Introduction} \label{sec:Intro}

Understanding the deconfining phase transition of 
QCD is a long-standing problem in high energy physics.
The situation is somewhat clearer in pure SU(N) Yang-Mills theory,
where it is understood that on length scales larger than $\sim1/T$
the relevant effective degrees of freedom are SU(N) spin variables 
(thermal Wilson loops) in three dimensions, obtained
by compactifying the Euclidean time direction.
In this picture,
the deconfining phase transition manifests as spontaneous
breaking of the Z(N) center symmetry 
\cite{'tHooft:1977hy,Polyakov:1978vu,Susskind:1979up,McLerran:1981pb}.


Numerical simulations show that the interaction measure, $(e-3p)/T^4$,
times $T^2/T_c^2$, is approximately flat from $\sim 1.2 T_c$ to $\sim
4 T_c$, for up to six colors
\cite{Umeda:2008bd,Panero:2009tv,Datta:2010sq,Borsanyi:2012ve}.
Similar behavior is observed in three space-time dimensions, for
$(e-2p)/T^3$ times $T/T_c$ \cite{Caselle:2011mn,Lucini:2012gg}.  This
shows that the leading corrections to an "ideal" gas term, $\sim T^4$
in four dimensions, and $\sim T^3$ in three dimensions, are terms
$\sim T^2$, in both four and three dimensions.  This differs from
corrections due to a bag pressure as in the MIT bag model, which would
be independent of temperature.  This behavior can be reproduced within
matrix models of the thermal Wilson
loop~\cite{Dumitru:2004gd,Dumitru:2003hp,Meisinger:2001cq,Meisinger:2001fi,Pisarski:2006hz,Dumitru:2010mj,Dumitru:2012fw}
with coupling constants which are rather simple functions of $T$.

To understand this phenomenon better, it is useful to study the
dynamics of the eigenvalues of the Wilson loops in different
representations of the gauge group.  The purpose of this work is to
study the effective potential of bare Wilson loops and of their
eigenvalues, in the fundamental and adjoint representations of SU(2),
through ab-initio simulations in discretized space-time. Our goal is
to shed some light on the validity of the perturbative ansatzes
for the effective action, which are used in the construction
of effective theories.  
Effective potentials for variables related to
the phases of the eigenvalues of thermal Wilson loops have also been
obtained from functional methods such as the Functional
Renormalization Group or Dyson-Schwinger equations typically using background
Landau-deWitt gauges \cite{Fister:2013bh,Braun:2007bx}, and our
results may be useful to benchmark these non-perturbative continuum
quantum field-theoretic computations. Also, we wish to verify that the
ansatz used to parametrize the results obtained in
Refs.\ \cite{Smith:2009kp,Smith:2010phd} for the classical potential of
a three dimensional matrix model of Wilson loops in the fundamental
representation is valid also for the full gauge theory.
 
\section{Setup}
We simulate pure SU(2) gauge theory using the standard
Wilson action
\be 
S= \beta \sum_\Box ( 1-\frac{1}{2} \rm {Re\, \Tr}~\boldu_\Box )~,
\label{Slatt}
\ee
where the sum runs over all plaquettes of the four dimensional
lattice. Lattice configurations are generated using an exact heatbath
algorithm \cite{Kennedy:1985nu} which generates link variables
according to the local Boltzmann distribution given by the staple
matrix. SU(2) matrices are stored in the quaternionic representation
\be
U=a_0 {\bf 1} + i \sum_{j=1}^3 a_j\, \sigma_j\quad, \sum_{j=0}^3 a_i^2=1~,\label{eq:quater1}
\ee
where $a_i \in \mathbb{R}$ and $\sigma_i$ are the Pauli matrices.  We
construct thermal Wilson loops $\boldl(\boldx)$ in the fundamental
representation in the usual way by multiplying temporal links:
\be
\boldl(\boldx)^{f}=\prod_{t=0}^{N_t-1}U_4(\boldx,t)~.\label{eq:ploop1}
\ee
Loops in higher representations can be generated from the fundamental
loop \cite{Gupta:2007ax}. We currently only also consider the adjoint
representation which is constructed as
\be
\boldl_{ij}^{a}=2\Tr \left(T_i \boldl^{f} T_j \boldl^{\dagger f} \right)~,\label{eq:adj1}
\ee
where $T_i$ are the three generators in the fundamental representation. 

We are interested in the per-site effective
potential $V_{\textrm{eff}}$ for the Polyakov loop $\ell \propto
\Tr\boldl $ and for the
eigenvalues of the Wilson line $\boldl$, in the fundamental or adjoint
representations. In other words, $V_{\textrm{eff}}(\ell)$ describes
fluctuations about the volume averaged mean field $\langle\ell\rangle$. 
The latter, in turn, determines the couplings $a(T)$, $b(T)$, $c(T)$ in
$V_{\textrm{eff}}(\ell)$, see below.

In our work, ``sites'' are understood as
the lattice-points of the three dimensional spin-system of the thermal
Wilson loops. Our approach differs from the computation 
of a potential for volume averaged fields
as in Refs.\ \cite{Fischer:2013eca,Langfeld:2013qia}.

We also comment on the difference of our approach to that of 
Ref.~\cite{Diakonov:2012dx}.  In this work they take
the Wilson line to be constant in time.  This is allowed, and
in the continuum, corresponds to a static gauge, $\partial_0 A_0 = 0$.
There is still a residual gauge freedom to fix spatial gauge transformations
at a given time.  For example, one could fix to Coulomb gauge, 
$\partial_i A_i = 0$.  Instead,
Ref.~\cite{Diakonov:2012dx} computes an effective potential by assuming
that the thermal Wilson line is constant in space.
In the continuum this corresponds
to setting $\partial_i A_0$ for all three spatial directions, and is
not an allowable gauge condition.
Because of this, their results are very different from ours or from 
Refs.~\cite{Fischer:2013eca,Langfeld:2013qia}.  In particular, 
\cite{Diakonov:2012dx} finds a potential which is completely flat in the
confined phase.

To obtain $V_{\textrm{eff}}$ we first compute the
per-site probability distribution $P$ by histogramming each
observable. From these distributions we obtain its classical or
``constraint'' effective potential $V_0=-\ln P$.  The effective
potential is then obtained from the Legendre transform of the
moment-generating function $W(h)$:
\be
W(h)= \ln \int dx\,\exp \left(-V_0(x)+hx \right)~. \label{eq:momgen}
\ee
Here, $x$ stands for the respective observable and the integral runs over the 
entire range of $x$. We obtain
\be
V_{\textrm{eff}}(\hat{x})= -\Gamma(\hat{x})~,
\ee
where
\be
\Gamma(\hat{x})=W(h(\hat{x}))-h(\hat{x})\hat{x} ~, ~~~ \hat{x}= \frac{dW(h)}{dh}~.
\label{eq:Legendre1}
\ee
In most cases considered here it is simpler to carry out the Legendre transformation via
\be
\Gamma(\hat{x})=\sup_h \left(\hat{x}h-W(h)\right)~.\label{eq:Legendre2}
\ee
This is our method of choice.\footnote{To be precise: in order to
  compute $\Gamma(\hat{x})$, for each data-set we take a proper
  parametrization of $V_0(x)$ as input. For a given $\hat{x}$ we
  change $h$ from $h=-100$ to $h=100$ in steps of $dh=0.02$.  For each
  $h$, $W(h)$ is computed from Eq.\ (\ref{eq:momgen}) using a $N$
  point Gaussian quadrature with $N=2048$. We obtain the maximum
  $\hat{x}h-W(h)$ with respect to $h$. We repeat this process for
  $400$ equidistant values of $\hat{x}$ in the domain of
  $\Gamma(\hat{x})$. We find that this procedure yields results which
  are stable under further increase of the resolution.}  We use
Eq.\ (\ref{eq:Legendre1}) only in some special cases where analytical
approximations to $W(h)$ can be easily obtained.

We use the string tension to fix the lattice spacing and thus the
temperature $T=1/(aN_t)$ in physical units. The exact procedure is
discussed at length in appendix \ref{sec:physunits} (in all figures,
the quoted $T/T_c$ are understood to have a relative uncertainty of
less than $\sim 2\%$). We change the temperature by varying $N_t$. The
advantage of this fixed scale approach is that it allows us to clearly
disentangle the effects of temperature changes and renormalization.

In the following we present a few formulae which shall be used
throughout this work.  Using Eqs.\ (\ref{eq:quater1}) and
(\ref{eq:adj1}) with $T_i=\sigma_i/2$ one can easily obtain the
compact expression 
\be \boldl_{ij}^{a}=2\left(a_i a_j + a_0
\sum_{k=1}^3 \epsilon_{ijk} a_k +
\delta_{ij}\left(a_0^2-\frac{1}{2}\right)\right)~,\label{eq:funadjoint_entries}
\ee 
which relates the elements of the adjoint matrices to the quaternionic
parameters $a_i$. From Eq.\ (\ref{eq:funadjoint_entries}) it
immediately follows that
\be
\Tr \boldl^{a}=|\Tr \boldl^{f}|^2 - 1~.\label{eq:funadjoint}
\ee
From (\ref{eq:quater1}) it follows that the eigenvalues $\lambda_{1,2}$ of the fundamental
loop are given by
\be
\lambda_{1,2}=a_0 \pm \sqrt{a_0^2-1}~.
\ee
The eigenvalues form a pair of complex conjugates which lie on the
unit circle. Also, they are related to the Polyakov loop $\ell$ by
\be
\ell^{f}=\frac{1}{2}\Tr \boldl^{f} = a_0 =\frac{1}{2}(\lambda_1 + \lambda_2)~.
\ee
Likewise, for the adjoint Wilson loop (which is a real 3-by-3 matrix)
one can easily show that the eigenvalues $\lambda_{1,2,3}$ are
\be
\lambda_1=1~,~\lambda_{2,3}=\frac{1}{2}(\Tr \boldl^{a} -1)\pm 
\sqrt{\frac{1}{4}(\Tr \boldl^{a} - 1)^2-1}~.
\ee
The non-unit eigenvalues $\lambda_{2,3}$ again form a pair of complex
conjugates with $|\lambda|=1$. Thus for both representations a single
phase $\phi\sim \ln \lambda$ uniquely fixes all eigenvalues.  The
$\boldl$ are related by a similarity transform to
$\mathrm{diag}(e^{i\phi},e^{-i\phi})$ and $\mathrm{diag}(1,e^{i\phi},e^{-i\phi})$
respectively, so that $\phi$ also fixes the trace of $\boldl$ through the
relations
\be
\Tr \boldl^{f}= 2 \cos(\phi^{f})~,\quad \Tr \boldl^{a}= 1+2 \cos(\phi^{a})~.\label{eq:wilsontrafo}
\ee
\section{Results}
We simulate $SU(2)$ gauge theory at $\beta=2.577856$ and
$\beta=2.635365$, which corresponds to $T_c$ at $N_t=10,12$.  The
lattice spacing is $a \sqrt{\sigma}=0.140$ and $a
\sqrt{\sigma}=0.116$, respectively.  We use time-like lattice sizes of
$N_t=12,10,8,6$.  For every lattice size and $\beta$ value several
hundreds of independent\footnote{Autocorrelations were investigated
  with the binning method \cite{berg2004:mc}.}  lattice
configurations were sampled from the equilibrium
distribution. Bin sizes for the histograms were chosen such that the
results appear as smooth lines at the resolutions of the figures
presented here. All distributions $P$ presented here are normalized.

There is a subtle issue regarding ergodicity of simulations in the
deconfined phase where the $Z(2)$ center symmetry is spontaneously
broken: In principle, at any finite lattice volume, an ergodic
algorithm which runs forever will tunnel infinitely often between the
two groundstates. Hence, all probability densities $P$ will display a
mixing of these groundstates and show no signs of a broken
symmetry. The physics presented here however is concerned with
the thermodynamic limit, for which the potential barrier between the
groundstates becomes infinitely large. Tunneling events should
therefore not occur, which strictly speaking violates ergodicity. From
a practical point of view it is justified to ignore this problem as
long as the lattice volumes considered are sufficiently large so
that tunneling is strongly suppressed. One should take care however to
ensure that this is truly the case for a given temperature, since such
tunneling events occur more frequently as one approaches $T_c$ from
above. One can achieve this, for instance, by monitoring the volume
average of the Polyakov loop and looking for sign flips.

The results presented here are unaffected by this issue.  The lowest
temperatures in the deconfined phase which we consider here are well
above $T_c$, so that even on a $32^3\times 4$ lattice tunneling is
practically impossible and the infinite volume limit is well
approximated. For each dataset discussed in this section simulations
were done on lattices with $N_x=48$ and $N_x=64$. We find that both
cases give identical results (indistinguishable histograms), which
confirms that finite-volume effects play no role. Likewise, this
confirms that statistical errors are negligible given the resolutions of
the presented figures. Without restricting generality we chose to
represent the symmetry broken phase as the state with
$\langle\ell\rangle >0$.

\subsection{The fundamental representation}

\begin{figure}[h]
\includegraphics*[width=\linewidth]{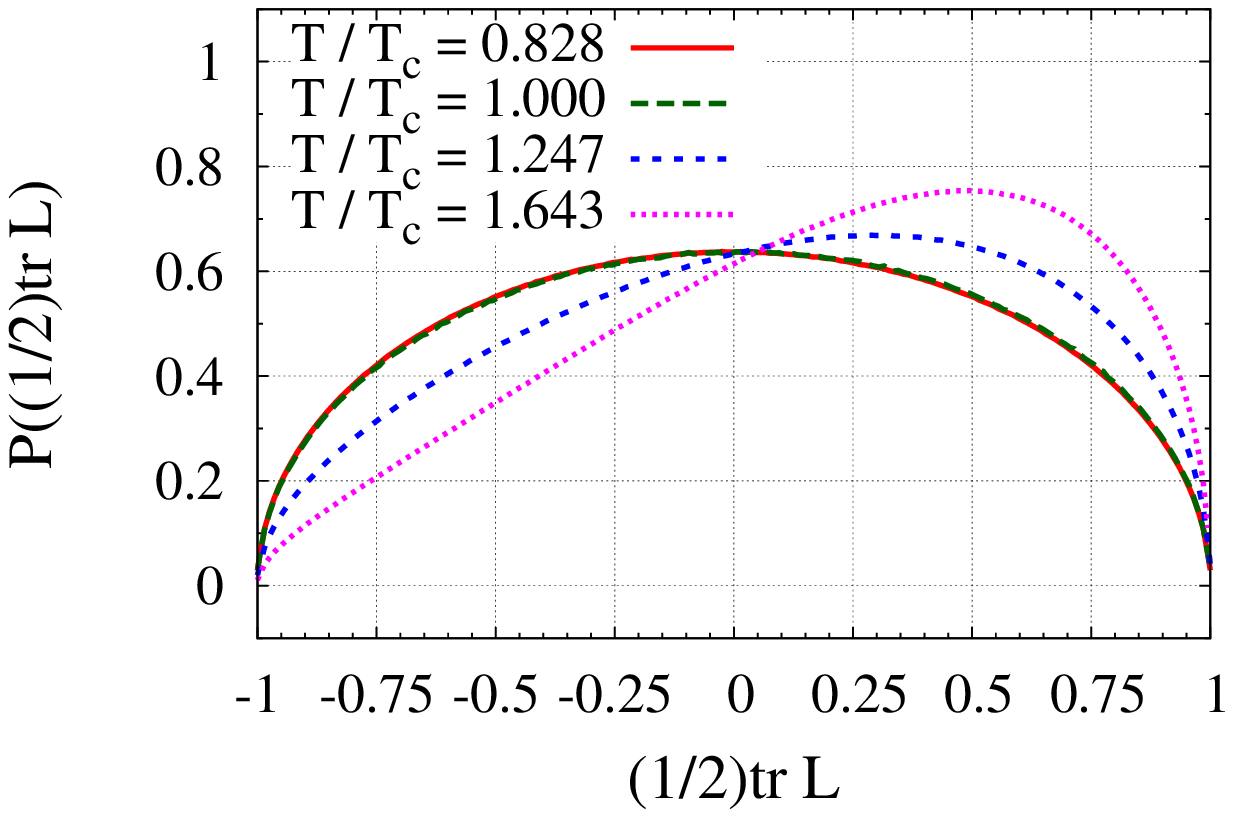}
\includegraphics*[width=\linewidth]{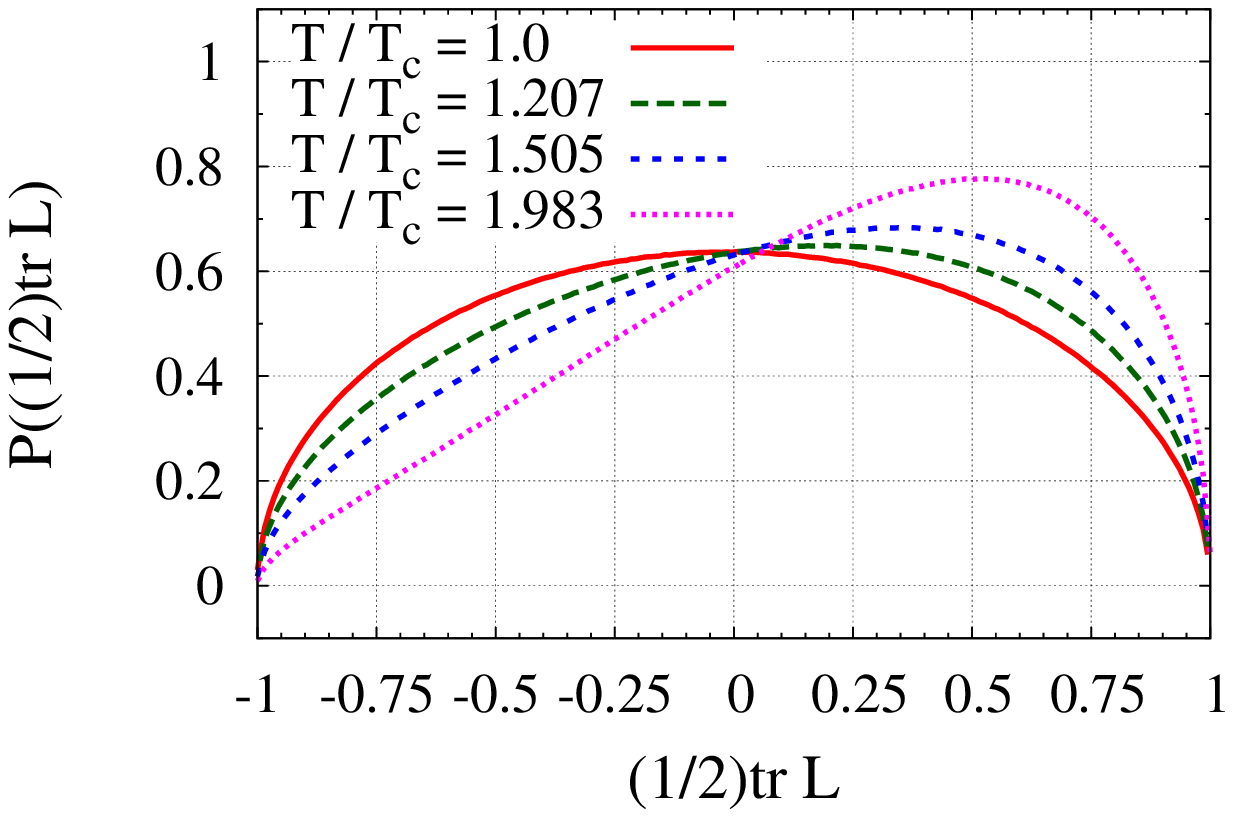}
\caption{Distribution of the SU(2) Polyakov loop in the fundamental
  representation at $\beta=2.577856$ (top) and $\beta=2.635365$
  (bottom).  The different temperatures correspond to $N_t=12,10,8,6$
  respectively.
\label{fig:su2fun1}}
\end{figure}

\begin{figure}[h]
\includegraphics*[width=\linewidth]{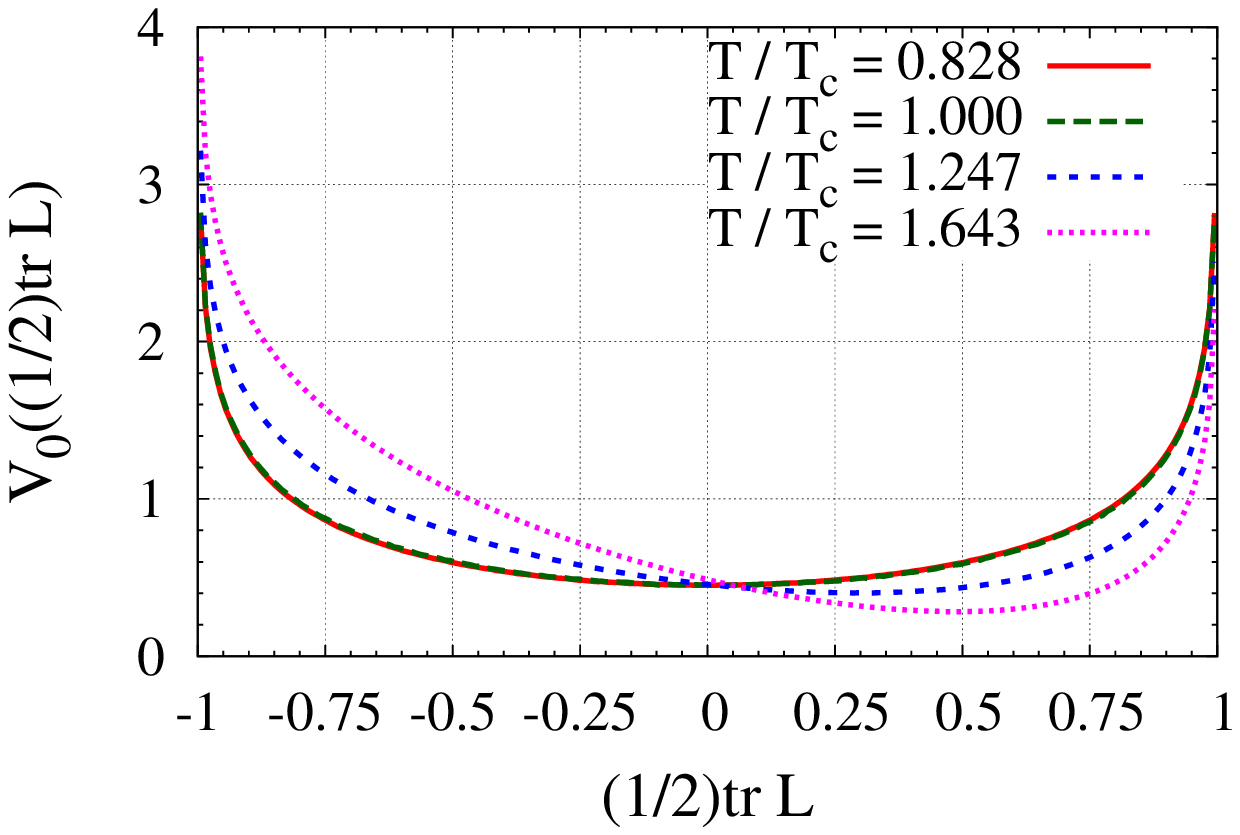}
\includegraphics*[width=\linewidth]{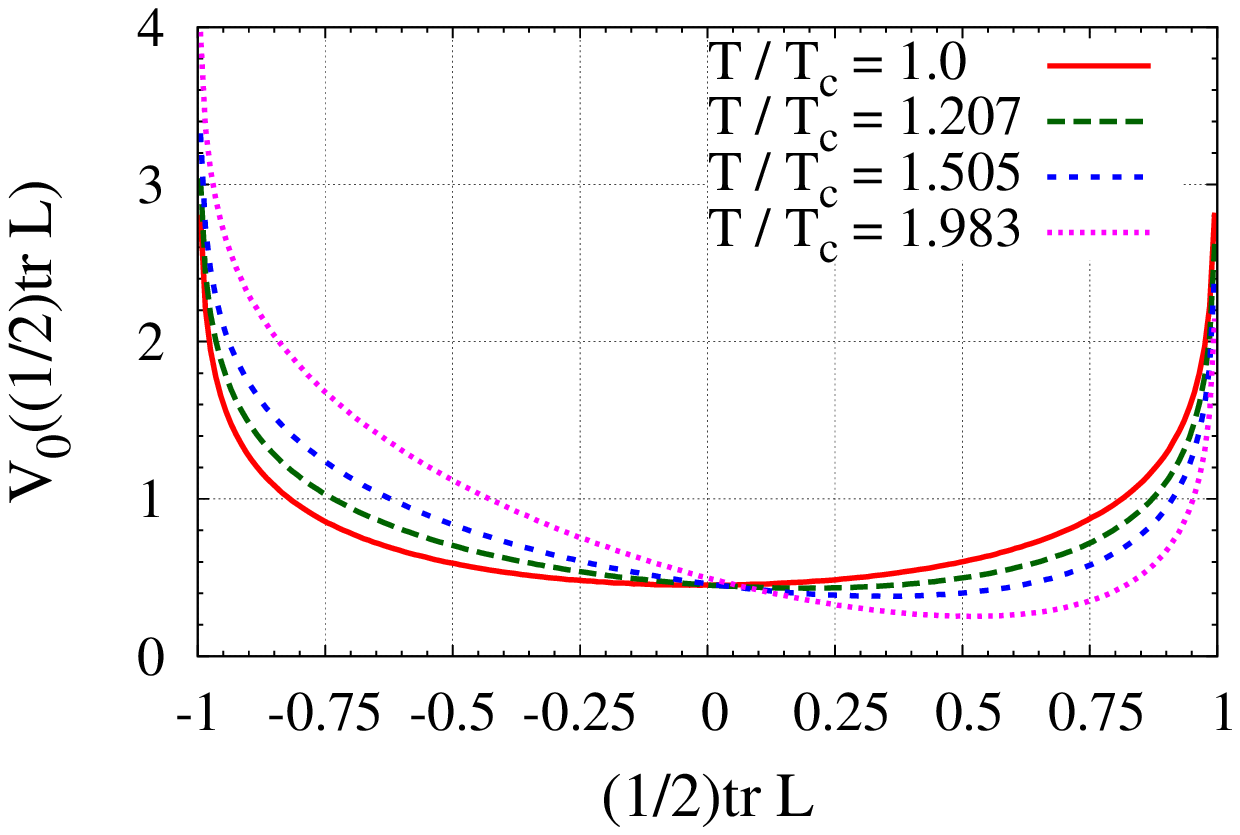}
\caption{Constraint effective potential of SU(2) Polyakov loop in
  fundamental representation at $\beta=2.577856$ (top) and
  $\beta=2.635365$ (bottom). The different temperatures correspond to
  $N_t=12,10,8,6$ respectively.
\label{fig:su2fun2}}
\end{figure}

We begin with the probability density $P(\ell)$ of
the Polyakov loop $\ell=(1/2)\Tr\boldl$ in the fundamental
representation. Fig.\ \ref{fig:su2fun1} shows $P(\ell)$
for  $\beta=2.577856$ and $\beta=2.635365$ at
$N_t=12,10,8,6$, which corresponds to temperatures
ranging from slightly below $T_c$ to roughly $2 T_c$
(smaller $\beta$ corresponds to lower $T$ for fixed $N_t$).
One can clearly see the effect of spontaneous breaking of the $Z(2)$
($\ell \to -\ell$) symmetry: right at $T_c$ the distributions are
symmetric around $\ell=0$.  In fact, we have confirmed that they are
identical for any $N_t \in [6,\ldots,12]$ at the respective
$\beta_c$. At higher temperature $P(\ell)$ is skewed and it develops a
peak at non-zero $\ell$ which moves towards $\ell=1$ as one raises the
temperature. In Fig.\ \ref{fig:su2fun2} one can see that the
corresponding constraint effective potential $V_0(\ell)=-\ln P(\ell)$
gets tilted, which is exactly what one expects at a second order
phase transition.

What is striking is that the lattice spacing affects the
distribution. Comparing the simulations at the two different values of
the coupling, $\beta=2.577856$ (corresponding to a larger lattice
spacing) clearly approaches the perturbative vacuum ($\langle |\ell|
\rangle \sim 1$) much faster with increasing temperature. It is well
known that the expectation value of the bare Polyakov loop is scale
dependent and vanishes in the continuum limit (hence the need to
construct renormalized loops
\cite{Gupta:2007ax,Dumitru:2003hp}). However, here we show that in
fact the entire distribution is affected in a profound way. In the
continuum $N_t\to\infty$ limit $P$ is deformed into the distribution
at $T=T_c$.  One way of thinking of this is that the $T=T_c$
distribution is an attractive ultra-violet fixed point of
renormalization group transformations. What is also noteworthy is that
the distribution appears not to change at $T<T_c$.\footnote{We do not
  rule out that there are small changes to the distribution below
  $T_c$ which we are not sensitive to. Such would be consistent with
  the findings of other authors
  \cite{Greensite:2012dy,Greensite:2013}. }

To make these statements quantitative we model the 
distributions using appropriate parametrizations. It turns out
that the $T=T_c$ distribution is exactly
\be
P^{(T_c)}(\ell)=\frac{2}{\pi}\sqrt{1-\ell^2}~,\label{eq:Vand1}
\ee
which corresponds to a random walk on the SU(2) group manifold
with the appropriate measure. The corresponding constraint potential
is the Vandermonde potential \cite{Dumitru:2004gd}
\be
V^{(T_c)}_0(\ell)=-\frac{1}{2}\ln\left(1-\ell^2 \right) -\ln\left(\frac{2}{\pi}\right)~.\label{eq:Vand2}
\ee
At $T>T_c$, for the entire range of temperatures considered here, the following
ansatz reproduces the simulated curves accurately:
\be
V_0(\ell)=V^{(T_c)}_0(\ell) + a(T)-b(T)\ell+c(T)\ell^2 ~.\label{eq:Vand3}
\ee
This $V_0(\ell)$ corresponds to the potential for a single ``local
spin'' $\ell(x)$ given that $\langle\ell\rangle \ge0$. If instead one
adopts the convention that $\langle\ell\rangle \le0$ by performing a
Z(2) transformation then $V_0(\ell)$ is obtained from the above by
letting $b\to-b$.

\begin{figure}[h]
\includegraphics*[width=\linewidth]{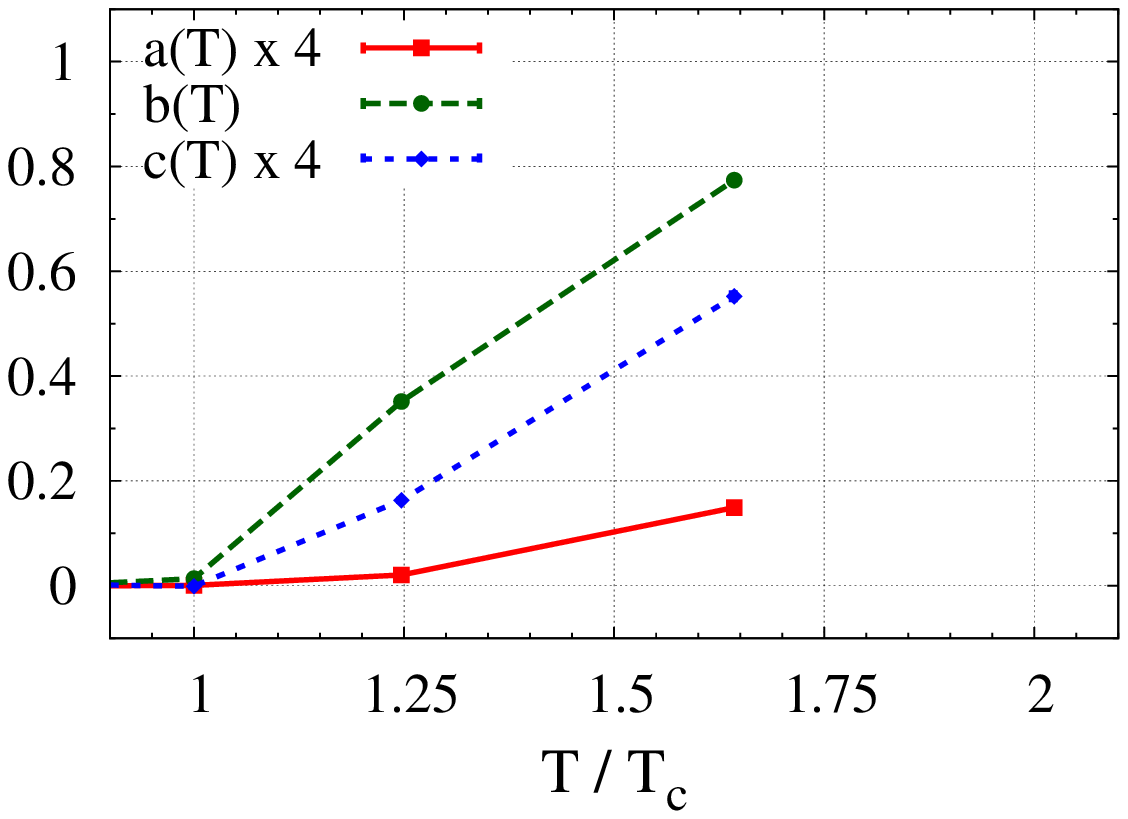}
\includegraphics*[width=\linewidth]{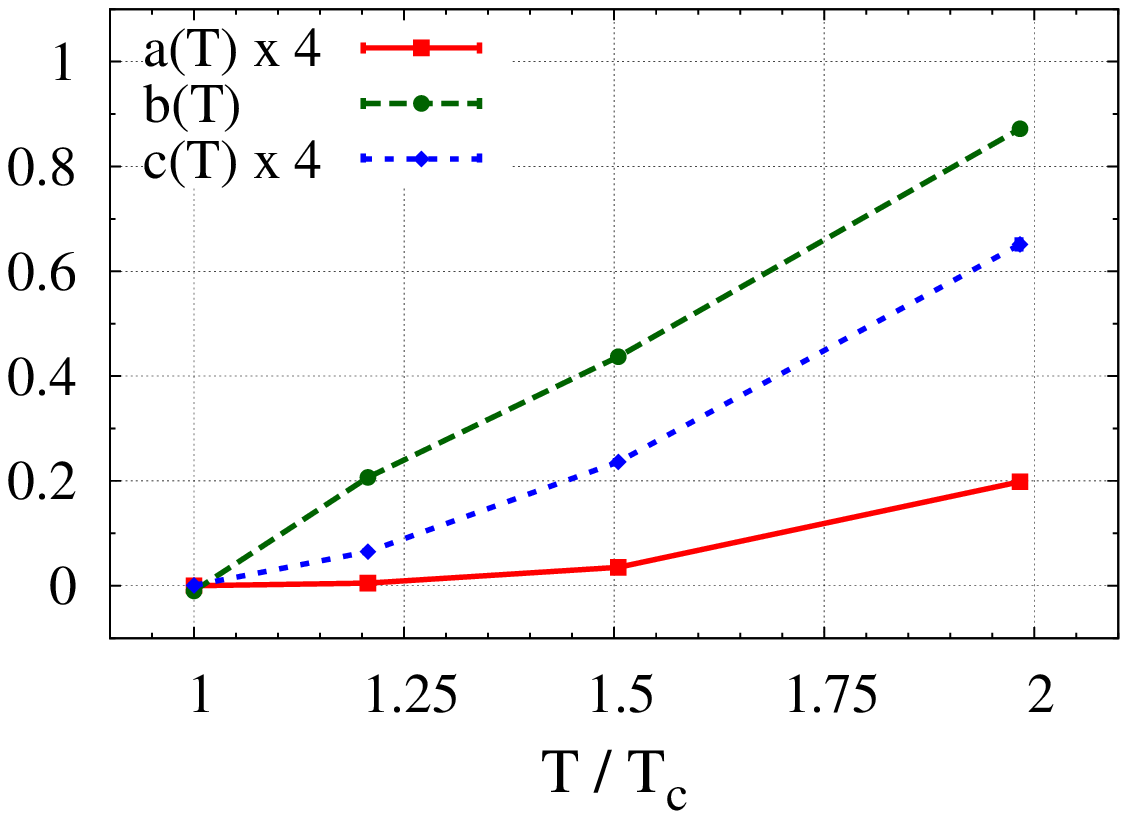}
\caption{Fit coefficients for $\beta=2.577856$ (top) and
  $\beta=2.635365$ (bottom).
$a(T)$ and $c(T)$ are scaled up by factors of $4$. The connecting lines
are there to guide the eye only but do not represent actual models. 
\label{fig:su2evfun5}}
\end{figure}

The potential~(\ref{eq:Vand3}) corresponds to the distribution
\be
P(\ell)=\frac{2}{\pi}\sqrt{1-\ell^2}\,\exp\left(-a(T)+b(T)\ell-c(T)\ell^2\right)
~. \label{eq:Vand4}
\ee
We obtain the parameters $a,b,c$  from a $\chi^2$ fit. We do not plot
the fits
as they are indistinguishable from the simulation results.
The temperature dependence of the parameters
for both values of $\beta$ are shown in Fig.\ \ref{fig:su2evfun5}. 
It appears that the term linear in $\ell$ is dominant for the
temperature range investigated here. It is also precisely this term
which drives breaking of the $Z(2)$ symmetry (it in some sense
describes the interaction of a particular site with the external field
generated by the symmetry broken state of the lattice as a whole).
The ansatz (\ref{eq:Vand3}) is exactly of the form which was used in
Refs.\ \cite{Smith:2009kp,Smith:2010phd} to parametrize the constraint
effective potential for $\ell$ in a $3D$ effective matrix theory of
$SU(2)$ Wilson lines, with a similar behavior of the parameters
$a,b,c$.
\begin{figure}[h]
\includegraphics*[width=\linewidth]{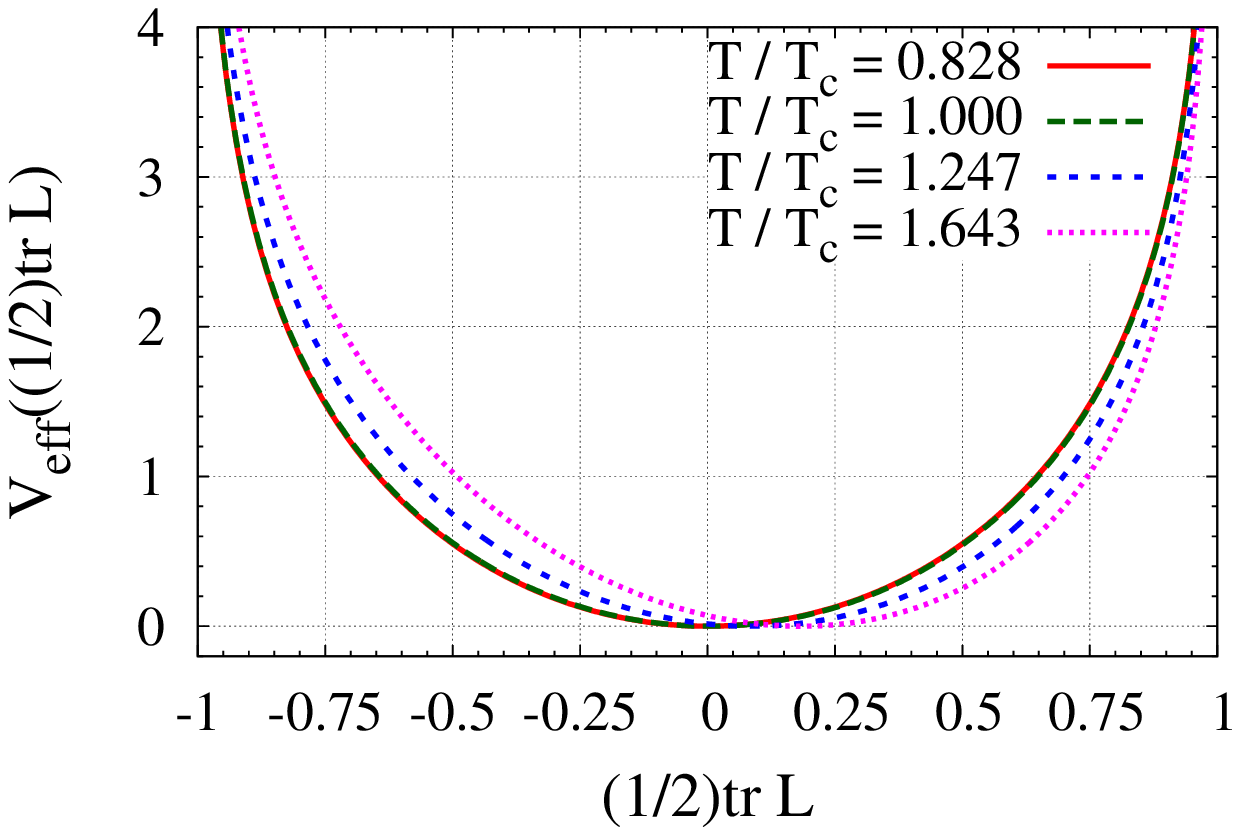}
\includegraphics*[width=\linewidth]{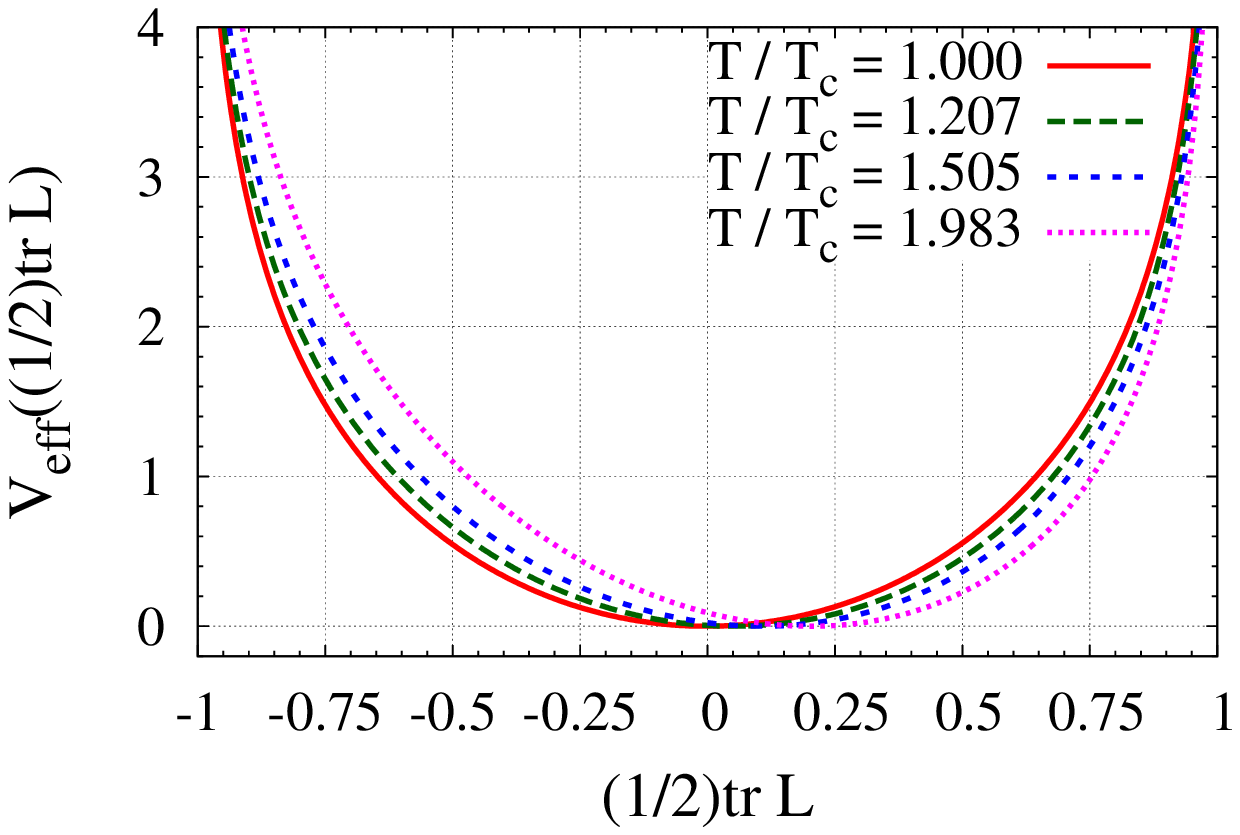}
\caption{Effective potential for the SU(2) Polyakov loop in the
  fundamental representation at $\beta=2.577856$ (top) and
  $\beta=2.635365$ (bottom). The quoted temperatures correspond to
  $N_t=12,10,8,6$ respectively.
\label{fig:su2fun2eff}}
\end{figure}

We obtain the effective potential $V_{\textrm{eff}}(\ell)$ by numerically
solving Eq.~(\ref{eq:Legendre2}), using the parametrization~(\ref{eq:Vand3})
of $V_0(\ell)$ as input. Obtaining the supremum numerically is
straightforward. The integral in Eq.~(\ref{eq:momgen}) is
computed numerically via Gaussian quadrature. The results
are shown in Fig.~\ref{fig:su2fun2eff}. We have confirmed
that the expectation value 
\be
\langle\ell\rangle=\int_{-1}^{1}d \ell\,\ell\,P(\ell)~,
\ee
coincides with the point at which $V_{\textrm{eff}}(\ell)$ is minimal,
as it should.

\begin{figure}[h]
\includegraphics*[width=\linewidth]{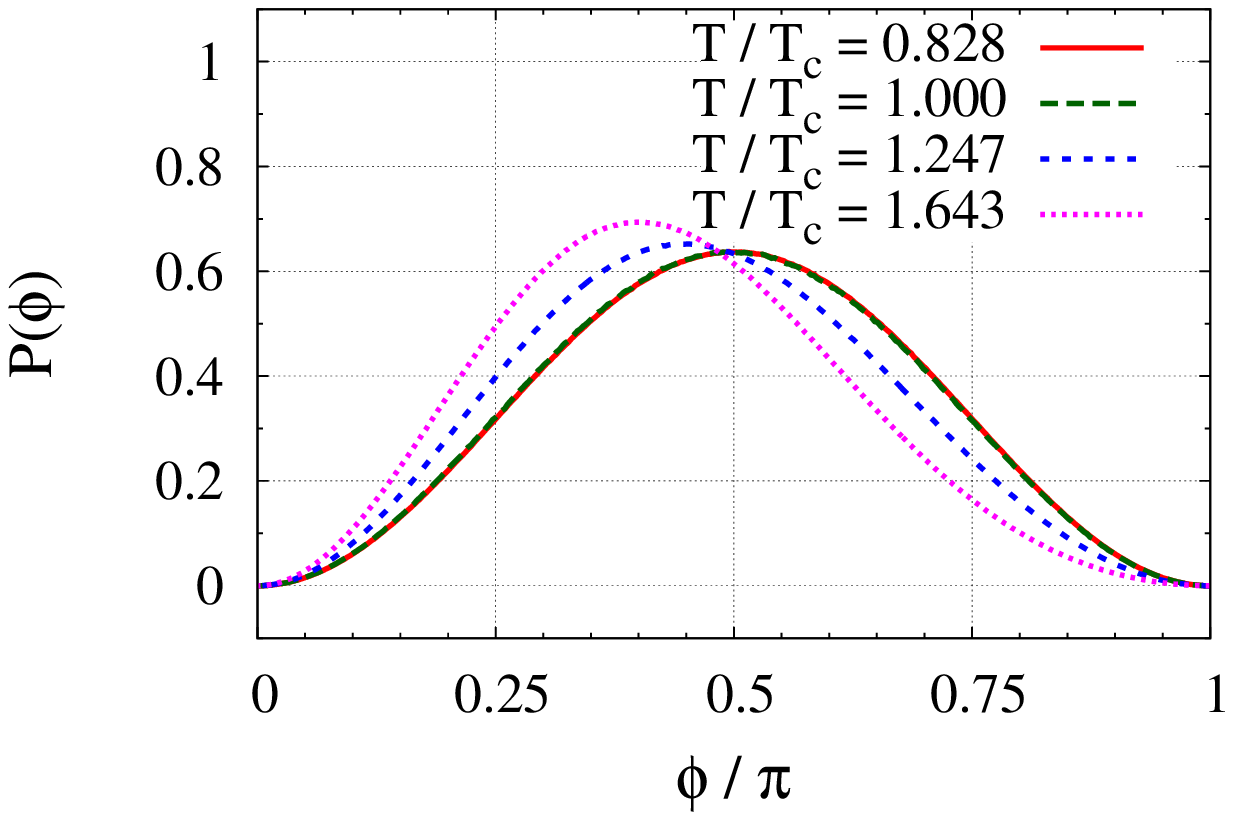}
\includegraphics*[width=\linewidth]{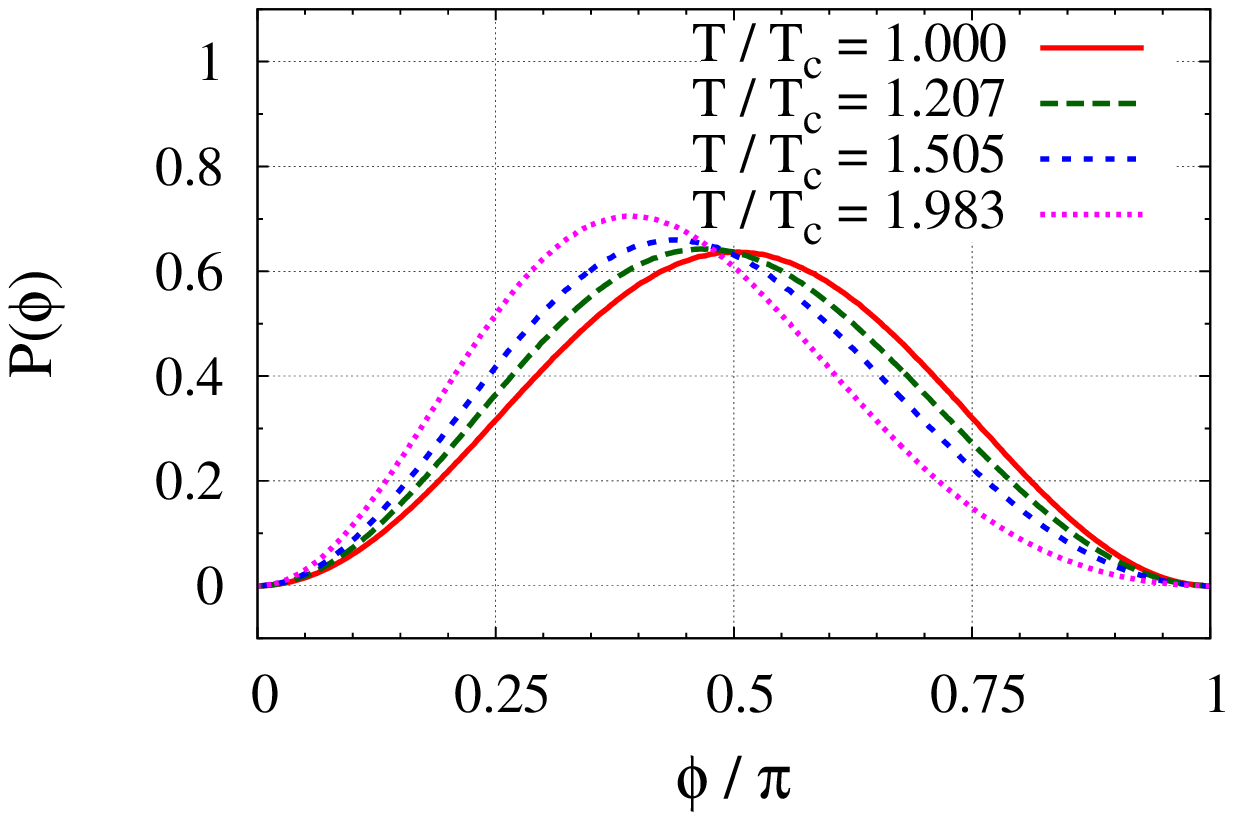}
\caption{Distribution of the phase of SU(2) Wilson loop eigenvalues in
  the fundamental representation at $\beta=2.577856$ (top) and
  $\beta=2.635365$ (bottom). The quoted temperatures correspond to
  $N_t=12,10,8,6$ respectively.
\label{fig:su2evfun3}}
\end{figure}

\begin{figure}[h]
\includegraphics*[width=\linewidth]{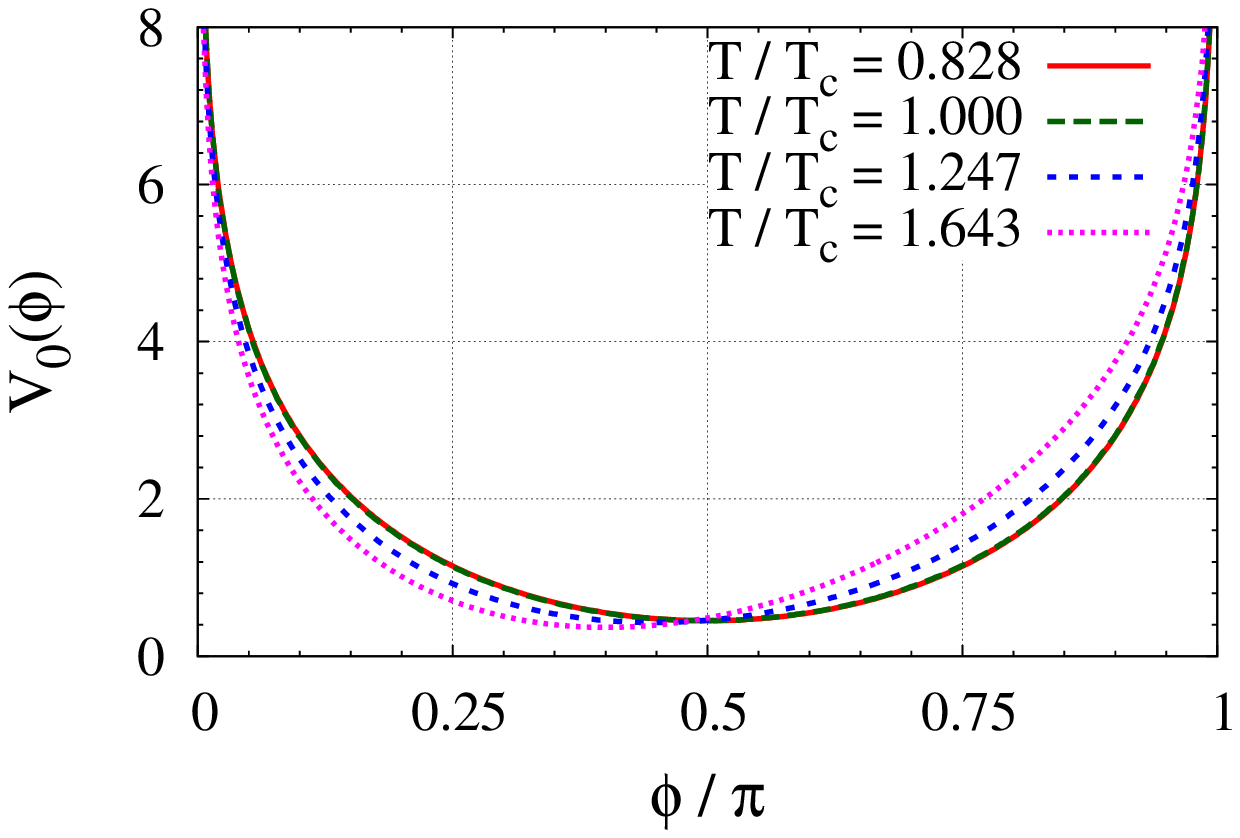}
\includegraphics*[width=\linewidth]{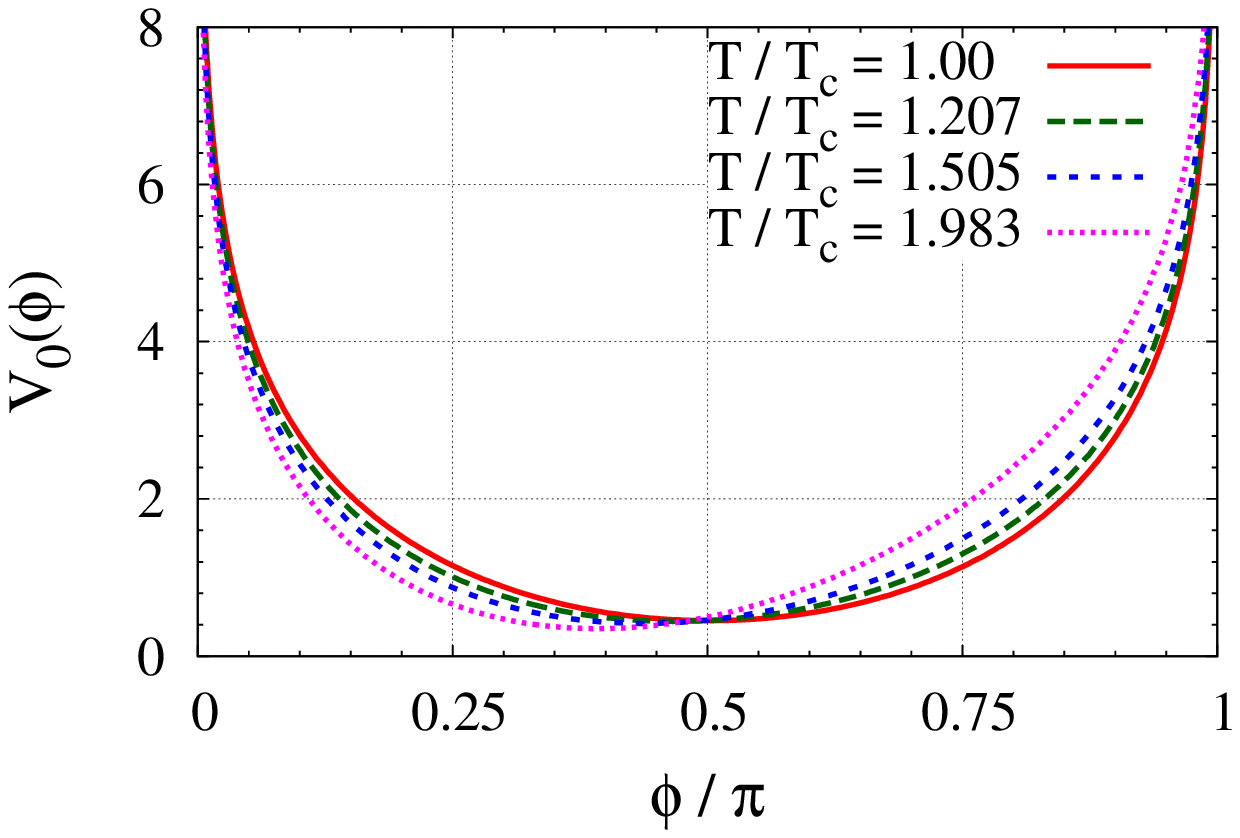}
\caption{Constraint effective potential for the phase of fundamental
  SU(2) Wilson loop eigenvalues at $\beta=2.577856$ (top) and
  $\beta=2.635365$ (bottom). The quoted temperatures correspond to
  $N_t=12,10,8,6$ respectively.
\label{fig:su2evfun4}}
\end{figure}
From the same ensembles we obtain the per-site distribution of the
phase $\phi$ of the eigenvalues of the Wilson loop $\boldl$.
Figs.\ \ref{fig:su2evfun3} and \ref{fig:su2evfun4} show the
distributions $P(\phi)$ and the potentials $V_0(\phi)$
respectively. One can see that right at $T_c$ the distribution is
peaked around $\boldl \sim \mathrm{diag}(i,-i)$, which is the state of
maximal eigenvalue repulsion. Increasing the temperature drives the peak
towards $\boldl \sim \mathrm{diag}(1,1)$ though eigenvalue
repulsion persists even above $T_c$. Such an effect is known to occur
also in matrix models of the Wilson line
\cite{Dumitru:2007ir,Smith:2008cm,Smith:2009kp}.

\begin{figure}[h]
\includegraphics*[width=\linewidth]{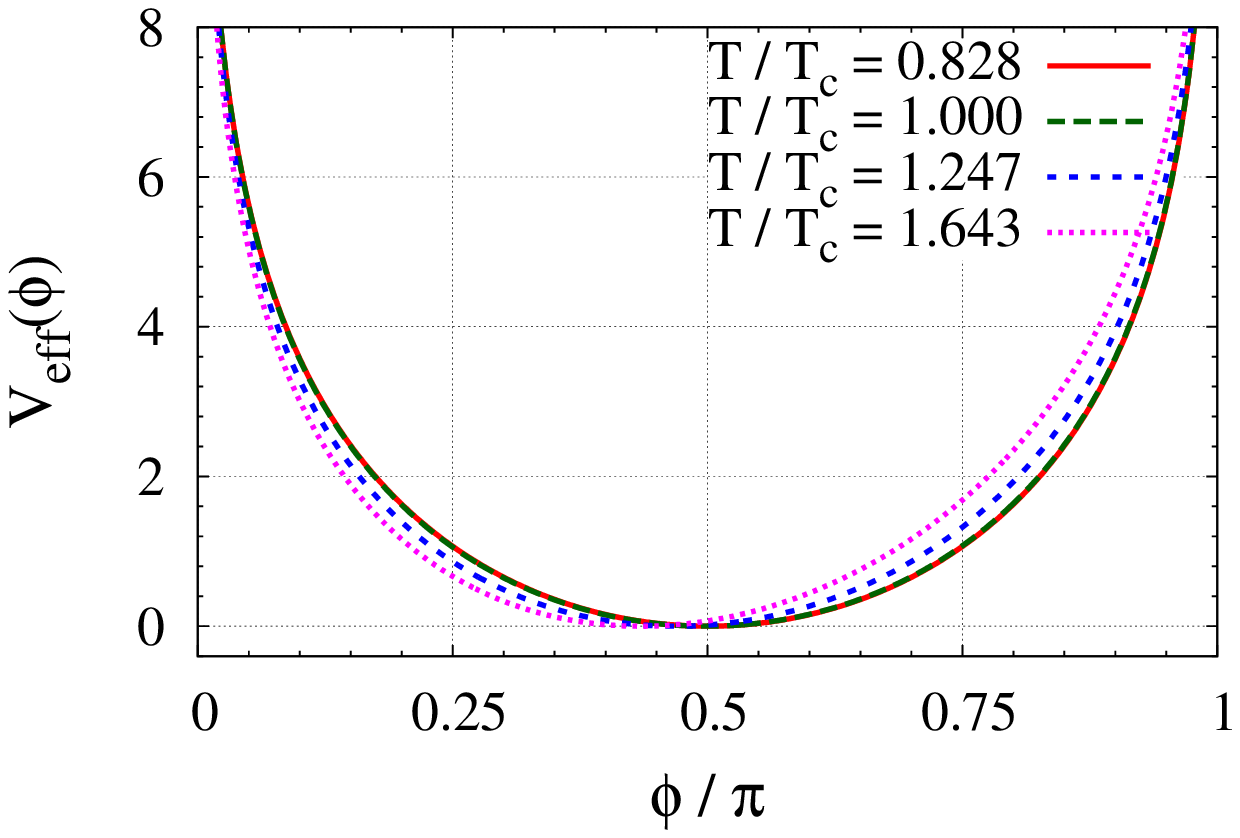}
\includegraphics*[width=\linewidth]{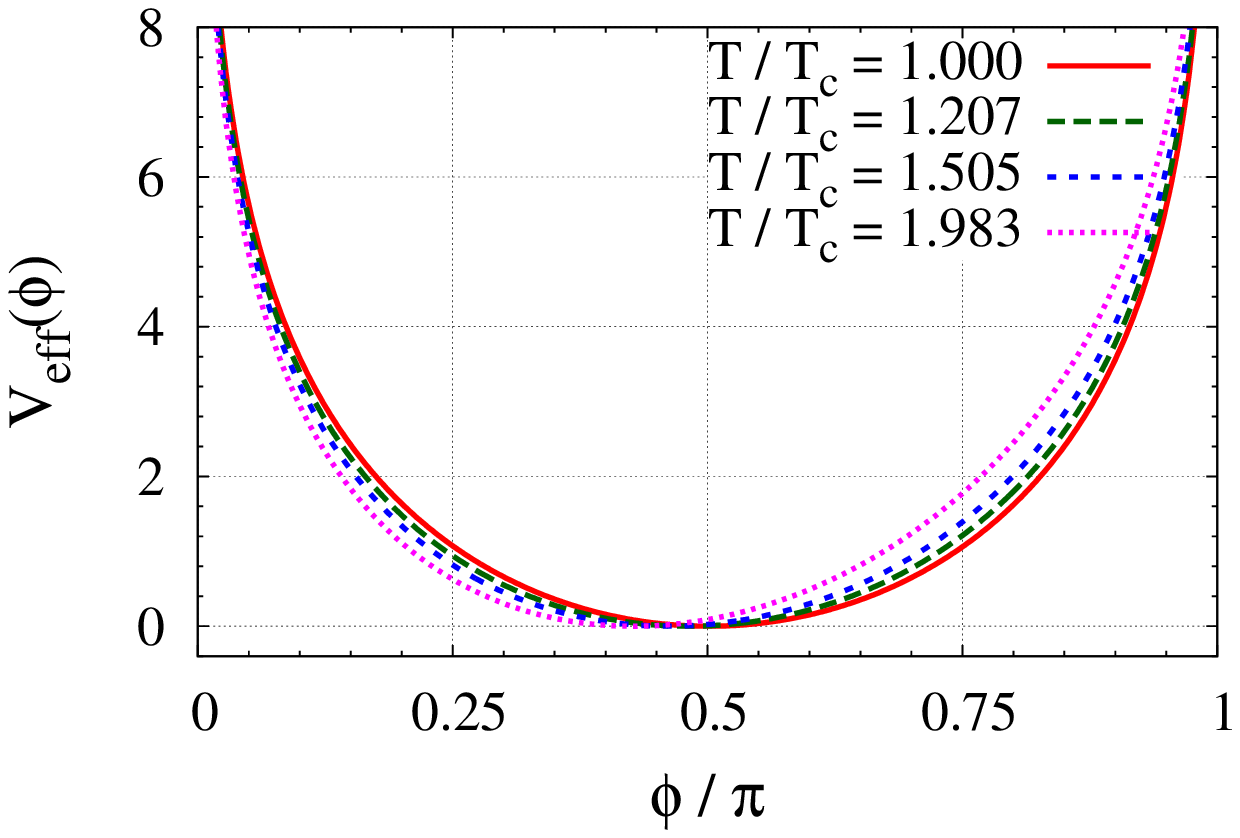}
\caption{Effective potential of phase of SU(2) Wilson loop eigenvalues in
fundamental representation at $\beta=2.577856$ (top) and
$\beta=2.635365$ (bottom).
The different temperatures correspond to $N_t=12,10,8,6$ respectively.
\label{fig:su2evfun4eff}}
\end{figure}

We can transform the parametrization of $P(\ell)$ given in
Eq.~(\ref{eq:Vand4}) to $P(\phi)$. Eq.~(\ref{eq:wilsontrafo}) provides
a one-to-one mapping from $\ell=[-1,\ldots,1]$ to
$\phi=[0,\ldots,\pi]$; we can thus use the standard integral
transformation law to obtain $P(\phi)$ from our models
(\ref{eq:Vand1}) and (\ref{eq:Vand4}). $\phi=[-\pi,\ldots,0]$ maps
into the same distribution $P(\ell)$, thus the restriction to positive
$\phi$ is legitimate. Using 
\be \int_{-1}^{1}d\ell\,P(\ell)
=\int_{\phi(\ell=-1)}^{\phi(\ell=1)} P(\ell(\phi))
\frac{d\ell}{d\phi}d\phi=\int_{\pi}^{0} P(\phi)d\phi~, 
\ee 
we arrive at
\be P^{(T_c)}(\phi)=\frac{2}{\pi}\sin^2{\phi}=\frac{1}{\pi}
\left(1+\cos(2\phi+\pi)\right)~, \label{eq:Vand5} 
\ee 
for the distribution right at $T_c$ and 
\bea
 P(\phi) &=& \frac{2}{\pi}\sin^2{\phi} \label{eq:Vand6}\\
& & \times \exp
\left(-a(T)+b(T)\cos(\phi)-c(T)\cos^2(\phi)\right)~,\nonumber
\eea
for the general case. The parameters $a,b,c$ are
unchanged. Eqs.~(\ref{eq:Vand5}) and (\ref{eq:Vand6}) again describe
the simulated results to such accuracy that the lines are
indistinguishable and thus the fit-curves are left out of the
figures. We again use our model function 
\bea
V_0(\phi)=&& \label{eq:Vand7}  \\
&& \hskip -1.2cm -\ln\left(\frac{2}{\pi}\sin^2{\phi}\right)
+a(T)-b(T)\cos(\phi)+c(T)\cos^2(\phi)~,\nonumber
\eea 
to obtain $V_{\textrm{eff}}(\phi)$. The Legendre transformation
(\ref{eq:Legendre2}) is carried out numerically, as before. The
results are shown in Fig.~\ref{fig:su2evfun4eff}.

In order to make our results portable (to compare them with other
calculations or use them as input for effective theories) we would
like to obtain analytical expressions for $V_{\textrm{eff}}(\phi)$ and
$V_{\textrm{eff}}(\ell)$. One possible approach is to again use
polynomial models (together with the Vandermonde contribution) and
adjust the parameters to fit the data. It turns out, however, that
this approach is not satisfactory as it requires high-order
polynomials and many parameters in both cases. Given how well our
models (\ref{eq:Vand4}) and (\ref{eq:Vand6}) described the simulated
results for $P(\ell)$ and $P(\phi)$, it would therefore be nice if the
Legendre transformation (\ref{eq:Legendre1}) could be carried out
exactly. Unfortunately this is not possible in closed
form. Nevertheless, an approximate solution can still be obtained
using the saddle point approximation. The next-to-leading order result
is easily obtained, and we will compare this result to our data in the
following.

\begin{figure}[h]
\includegraphics*[width=\linewidth]{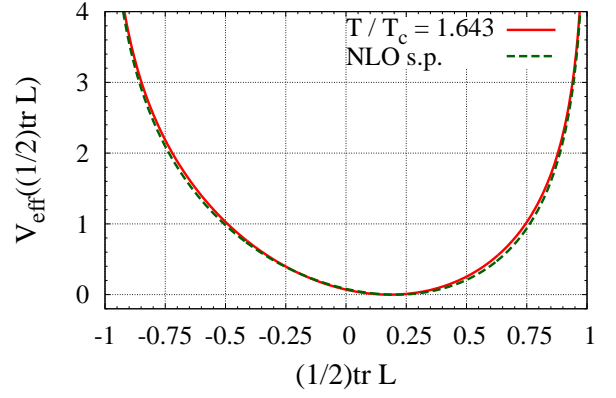}
\includegraphics*[width=\linewidth]{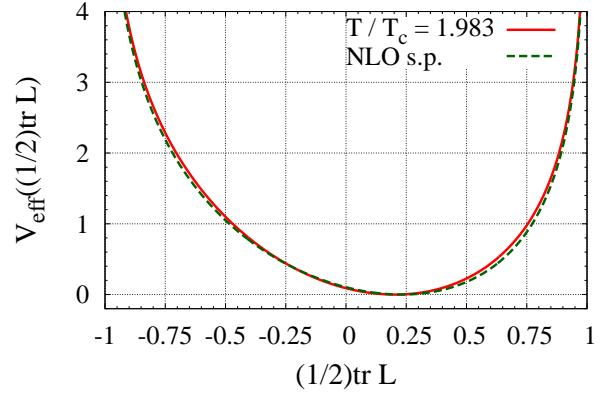}
\caption{Effective potential for SU(2) Polyakov loop in fundamental
  representation at $\beta=2.577856$ (top) and $\beta=2.635365$
  (bottom).  $N_t=6$ simulation result compared to NLO saddle point
  expansion. The SP curve has been shifted vertically to bring the minimum to
  zero.
\label{fig:su2fun2eff_spcompare}}
\end{figure}

\begin{figure}[h]
\includegraphics*[width=\linewidth]{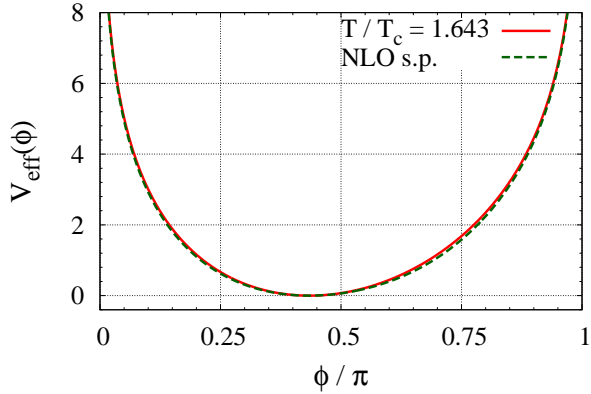}
\includegraphics*[width=\linewidth]{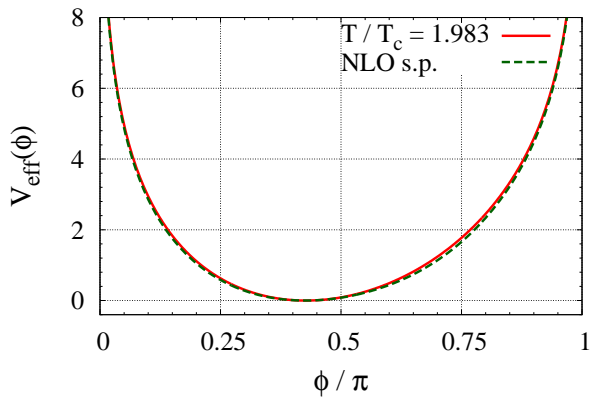}
\caption{Effective potential for the phase of SU(2) Wilson loop
  eigenvalues in the fundamental representation at $\beta=2.577856$
  (top) and $\beta=2.635365$ (bottom).  $N_t=6$ result compared to NLO
  saddle point expansion. The SP curve has been shifted vertically to bring the
  minimum to zero.
\label{fig:su2evfun4eff_spcompare}}
\end{figure}

Consider a stationary point $x_0$ of the exponential appearing in
Eq.\ (\ref{eq:momgen}) such that
\be
V_0'(x_0)=h~.\label{eq:saddle1}
\ee
We can expand the exponent in Eq.~(\ref{eq:momgen}) around this point.
Neglecting terms containing derivatives of higher than second
order we can carry out the integral:
\bea
W(h) &=& \ln \int_{-\infty}^{\infty}dx\,
\exp\left(-V_0(x)+hx\right) \\
&& \hskip -1.4cm \approx\ln \int_{-\infty}^{\infty} dx \,
\exp\big( -V_0(x_0)+hx_0 
-\frac{1}{2}V_0''(x_0)(x_0-x)^2 \big)\notag\\
&=&-V_0(x_0)+hx_0+\frac{1}{2}\ln
\left(\frac{V_0''(x_0)}{2\pi}\right)~. \notag
\eea
It holds that
\be
\frac{dW}{dh}=x_0 + \calo( V_0''' )~,
\ee
and thus the NLO effective potential is given by
\be
V_{\textrm{eff}}(x)\approx V_0(x)+\frac{1}{2}\ln
\left(\frac{V_0''(x)}{2\pi} \right)~.
\ee
The leading order is just the classical potential.  This formula can
be applied to the potentials for both $\ell$ and $\phi$.  From
Eqs.~(\ref{eq:Vand7}) and~(\ref{eq:Vand3}) we obtain the second
derivatives
\be
V_0''(\ell)= 2c(T)+\frac{1+\ell^2}{(1-\ell^2)^2}~,
\ee
and
\be
V_0''(\phi)= \frac{2}{\sin^2(\phi)}+b(T)\cos(\phi)+2
c(T)(\sin^2(\phi)-\cos^2(\phi))~.
\ee
In Figs.~\ref{fig:su2fun2eff_spcompare} and
\ref{fig:su2evfun4eff_spcompare} we compare the NLO saddle point
approximation to the exact results for both $V_{\textrm{eff}}(\ell)$
and $V_{\textrm{eff}}(\phi)$. The approximate curves are corrected by
an additive shift to account for the loss of normalization of $P$
which occurs. The shift is such that $V_{\textrm{eff}}$ is moved to zero at its minimum. 
In both cases $N_t=6$ is chosen as
an example. It appears that the approximation is quite good.

Lastly, we investigate how well the classical potentials $V_0(\ell)$
and $V_0(\phi)$ can be approximated by Taylor polynomials around the
confined ($\phi=\pm \pi/2$) or deconfined ($\phi=0,\pi$) vacua.
Expanding about $\phi=\pi/2$ we obtain
\bea
V_{0,1}(\phi)&\approx& -\ln\left(\frac{2}{\pi}\right) +a(T) +b(T) \left(\phi - \frac{\pi}{2}\right)\notag\\
&& + \left(c(T)+1\right)\left(\phi - \frac{\pi}{2}\right)^2
-\frac{b(T)}{6} \left(\phi - \frac{\pi}{2}\right)^3 \notag\\
&&+\left(\frac{1}{6}-\frac{c(T)}{3} \right)\left(\phi - \frac{\pi}{2}\right)^4  ~.\label{eq:Vand8}
\eea
Likewise, about $\phi=0$
\bea
V_{0,2}(\phi)&\approx& -\ln\left(\frac{2}{\pi}\sin^2(\phi) \right) +a(T) -b(T) +c(T)\notag\\
&& + \left(\frac{b(T)}{2}-c(T)\right) \phi^2\notag\\
&& +\left(\frac{c(T)}{3}-\frac{b(T)}{24} \right)\phi^4  ~.\label{eq:Vand9}
\eea
The logarithmic term has not been expanded since it diverges at
$\phi=0$.

Finally, an expansion around $\phi=\pi$ yields
\bea
V_{0,3}(\phi)&\approx& -\ln\left(\frac{2}{\pi}\sin^2(\phi) \right) +a(T) +b(T) +c(T)\notag\\
&& + \left(\frac{-b(T)}{2}-c(T)\right) \left(\phi-\pi\right)^2\notag\\
&& +\left(\frac{c(T)}{3}-\frac{-b(T)}{24}
\right)\left(\phi-\pi\right)^4  ~.\label{eq:Vand10}
\eea
Note that~(\ref{eq:Vand10}) is nothing but the Z(2) transform
of~(\ref{eq:Vand9}) corresponding to $\phi\to\pi-\phi$ and $b\to-b$.

\begin{figure}[h]
\includegraphics*[width=\linewidth]{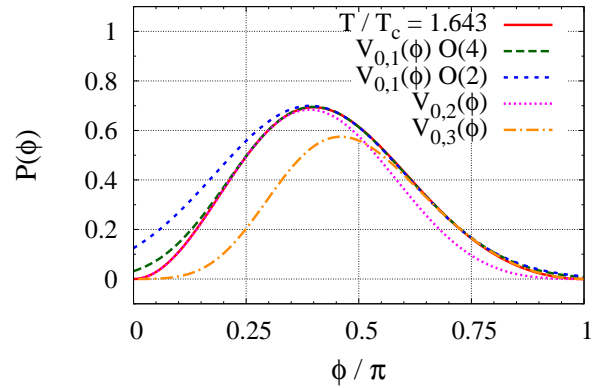}
\includegraphics*[width=\linewidth]{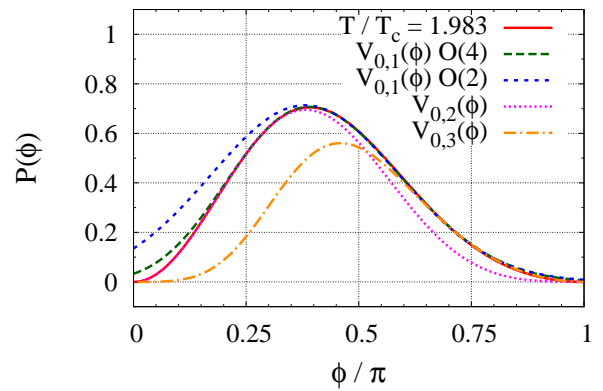}
\caption{Distribution of phase of SU(2) Wilson loop eigenvalues in
the fundamental representation at $\beta=2.577856$ (top) and
$\beta=2.635365$ (bottom). $N_t=6$ results compared to Taylor expansions.
\label{fig:su2evfun_taylor}}
\end{figure}

\begin{figure}[h]
\includegraphics*[width=\linewidth]{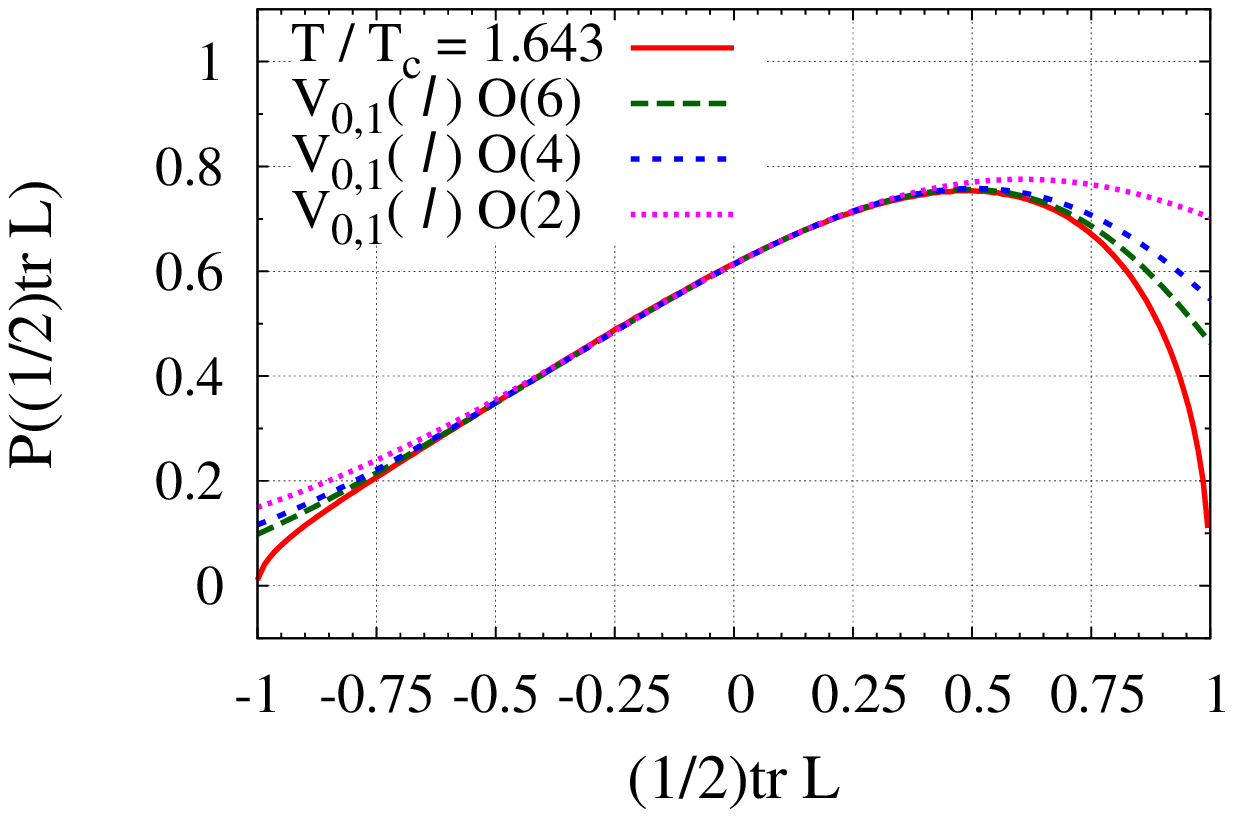}
\includegraphics*[width=\linewidth]{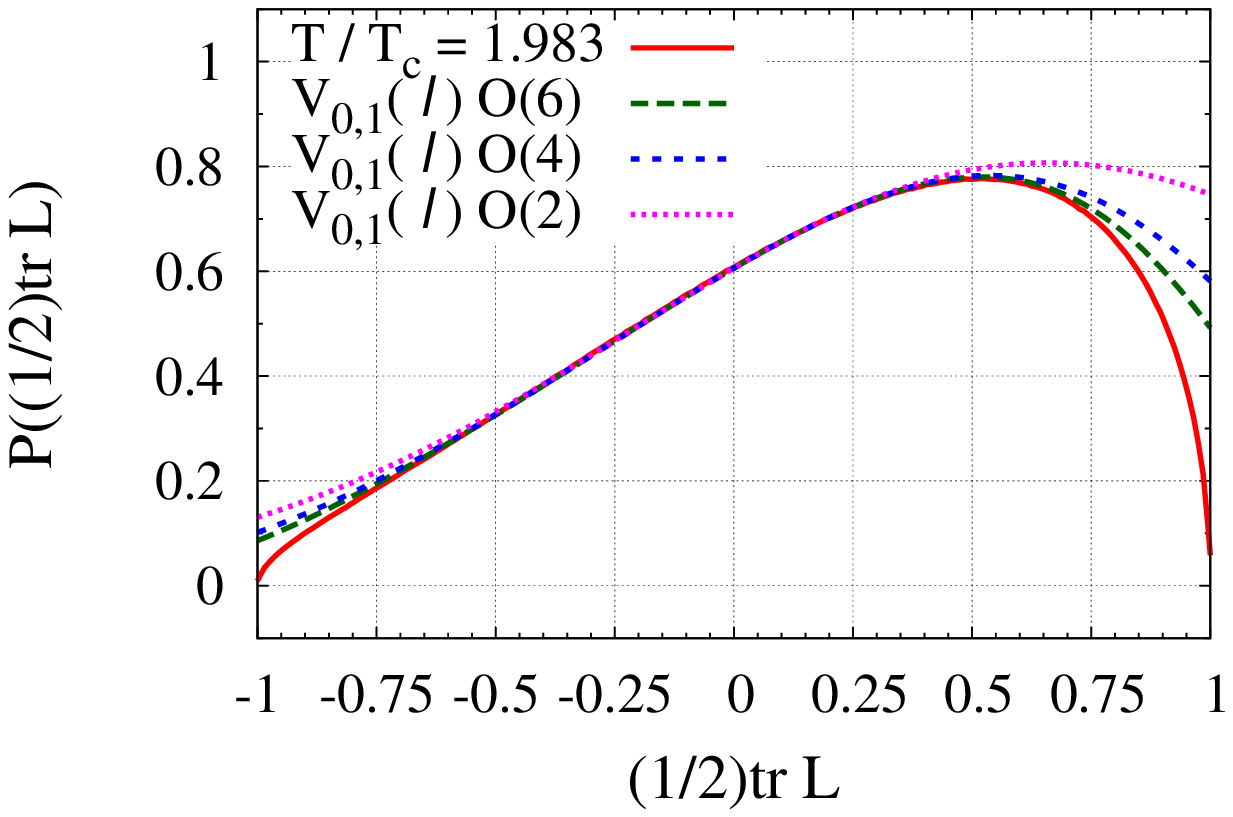}
\caption{Distribution of SU(2) Polyakov loop in
the fundamental representation at $\beta=2.577856$ (top) and
$\beta=2.635365$ (bottom).
$N_t=6$ results compared to Taylor expansions.
\label{fig:su2fun_taylor}}
\end{figure}

Fig.~\ref{fig:su2evfun_taylor} compares these expansions to the
simulation results for $N_t=6$, which corresponds to the highest
temperature for each respective $\beta$. It can be seen that
Eq.~(\ref{eq:Vand8}) works well, even if one terminates the expansion
at $\calo(2)$. The approximation smoothly approaches the simulation
results when at higher orders, and $\calo(4)$ reproduces the exact
result to high accuracy. Convergence is slower for
Eq.~(\ref{eq:Vand9}) and very slow for (\ref{eq:Vand10}). The latter
is not extremely suprising since $\phi=\pi$ is the wrong groundstate.
A similar expansion can be constructed for the
logarithmic contribution to $V_0(\ell)$. Expanding Eq.~(\ref{eq:Vand3})
to $\calo(6)$ about $\ell=0$ we get
\bea
V_{0,1}(\ell)&\approx& -\ln\left(\frac{\pi}{2}\right) +a(T)- b(T)\ell \notag\\
&& +\left(c(T)+\frac{1}{2}\right)\ell^2
+\frac{1}{4}\ell^4+\frac{1}{6}\ell^6
~.\label{eq:Vand11}
\eea
Fig.~\ref{fig:su2fun_taylor} shows the convergence of this expansion
towards the simulated curves for $N_t=6$. High orders are required
to accurately reproduce the data.

\subsection{The adjoint representation}

\begin{figure}[h]
\includegraphics*[width=\linewidth]{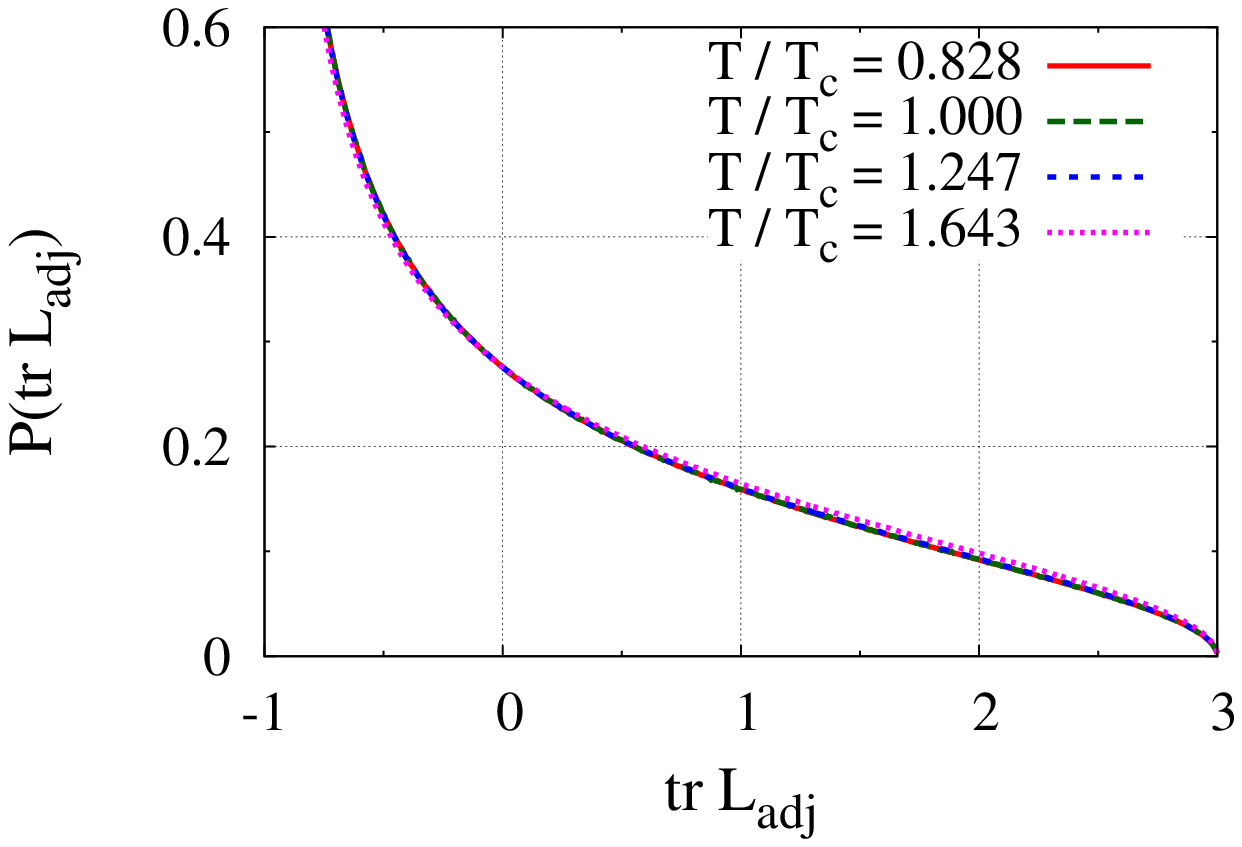}
\includegraphics*[width=\linewidth]{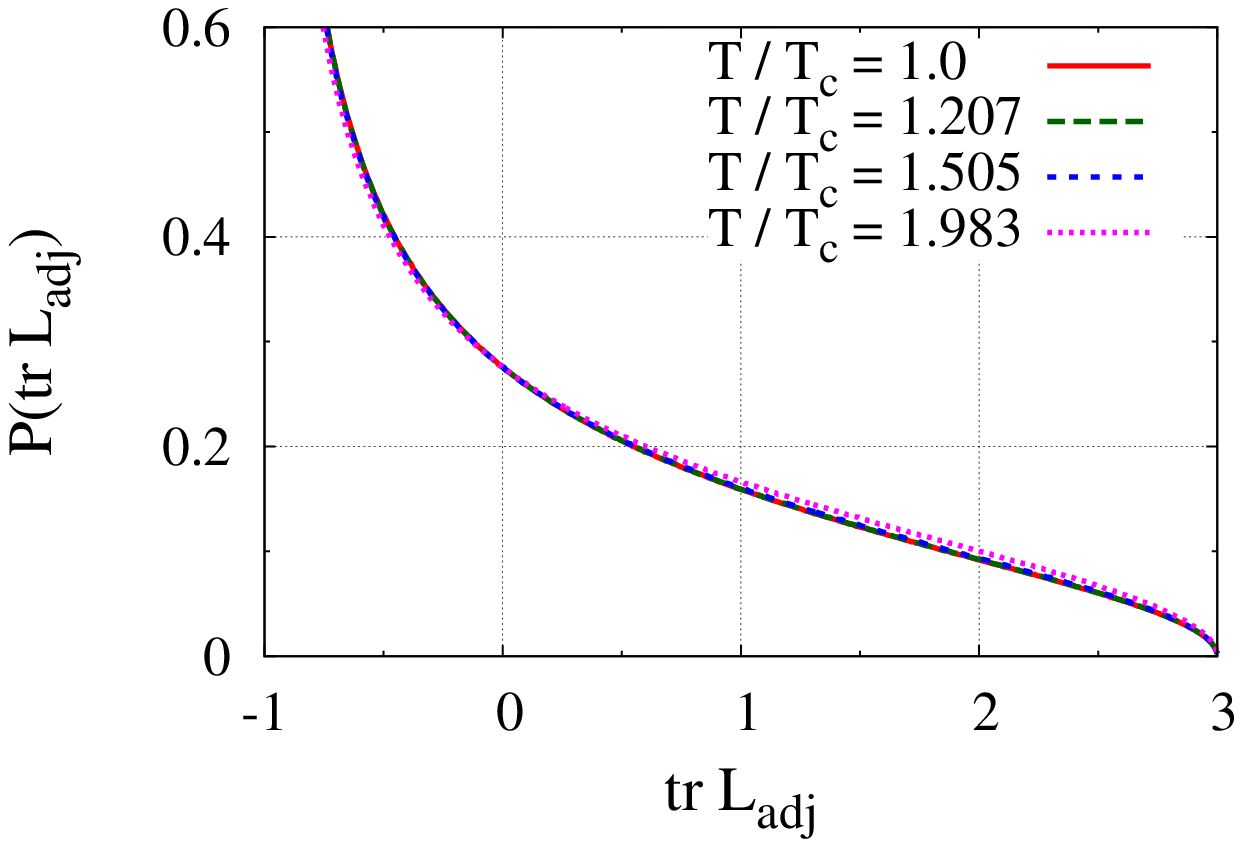}
\caption{Distribution of SU(2) Polyakov loop in adjoint representation
at $\beta=2.577856$ (top) and $\beta=2.635365$ (bottom). 
The different temperatures correspond to $N_t=12,10,8,6$ respectively.
\label{fig:su2adj1}}
\end{figure}

\begin{figure}[h]
\includegraphics*[width=\linewidth]{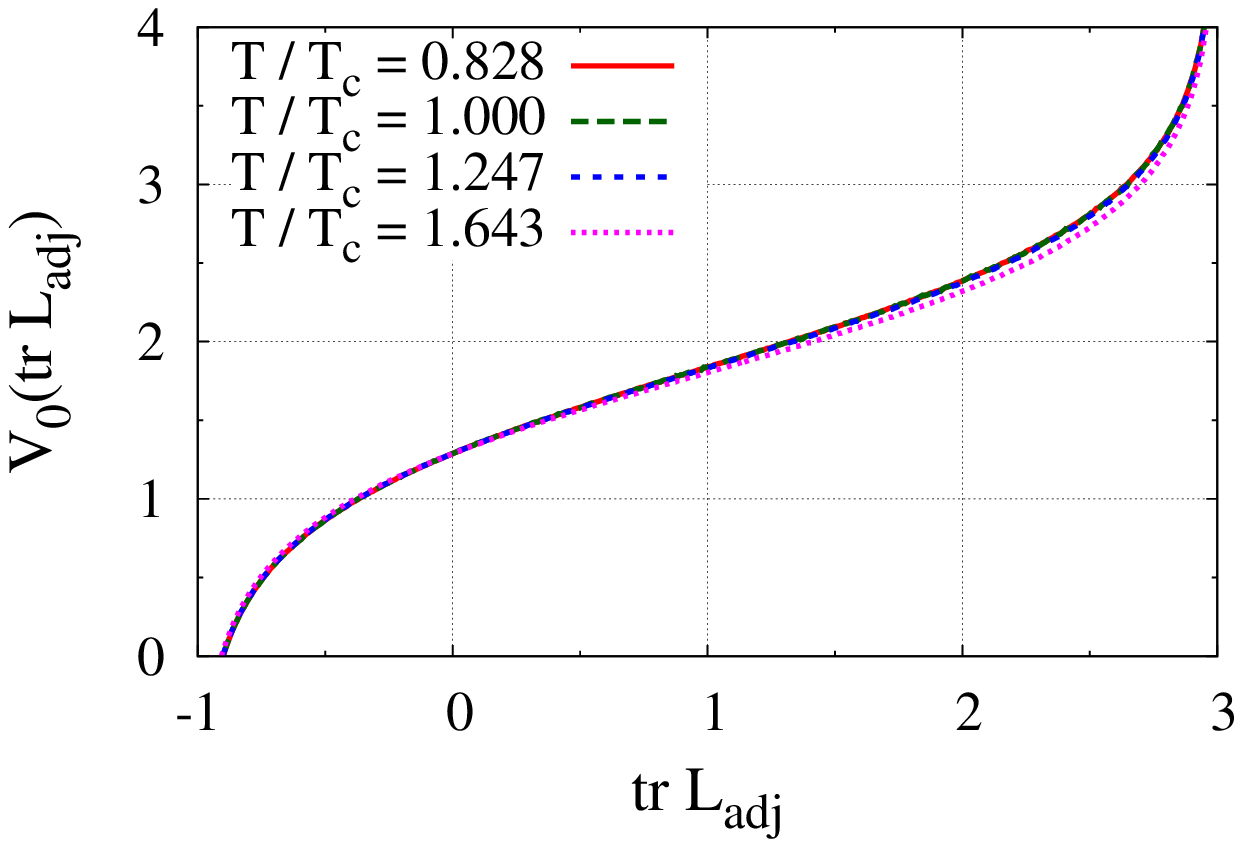}
\includegraphics*[width=\linewidth]{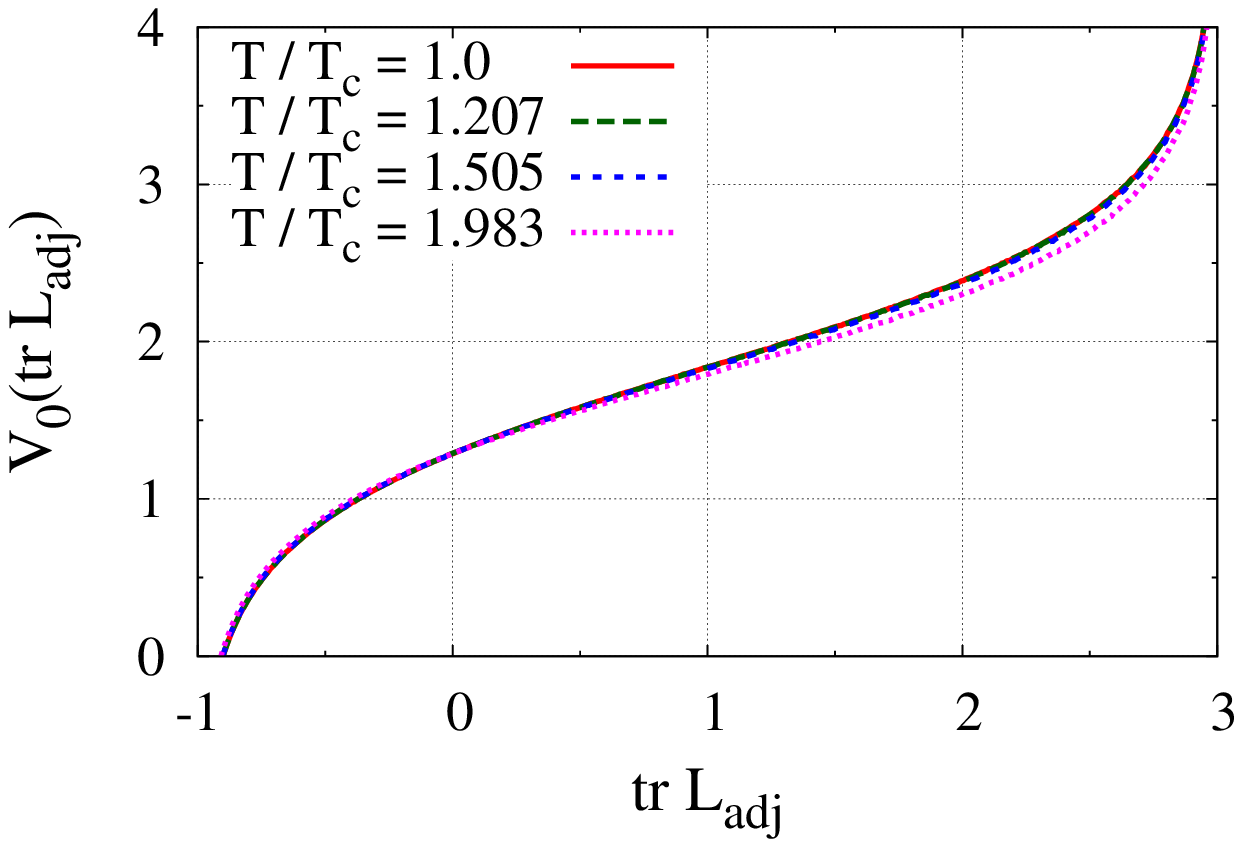}
\caption{Constraint effective potential of SU(2) Polyakov loop in
adjoint representation at $\beta=2.577856$ (top) and $\beta=2.635365$ (bottom). 
The different temperatures correspond to $N_t=12,10,8,6$ respectively.
\label{fig:su2adj2}}
\end{figure}

\begin{figure}[h]
\includegraphics*[width=\linewidth]{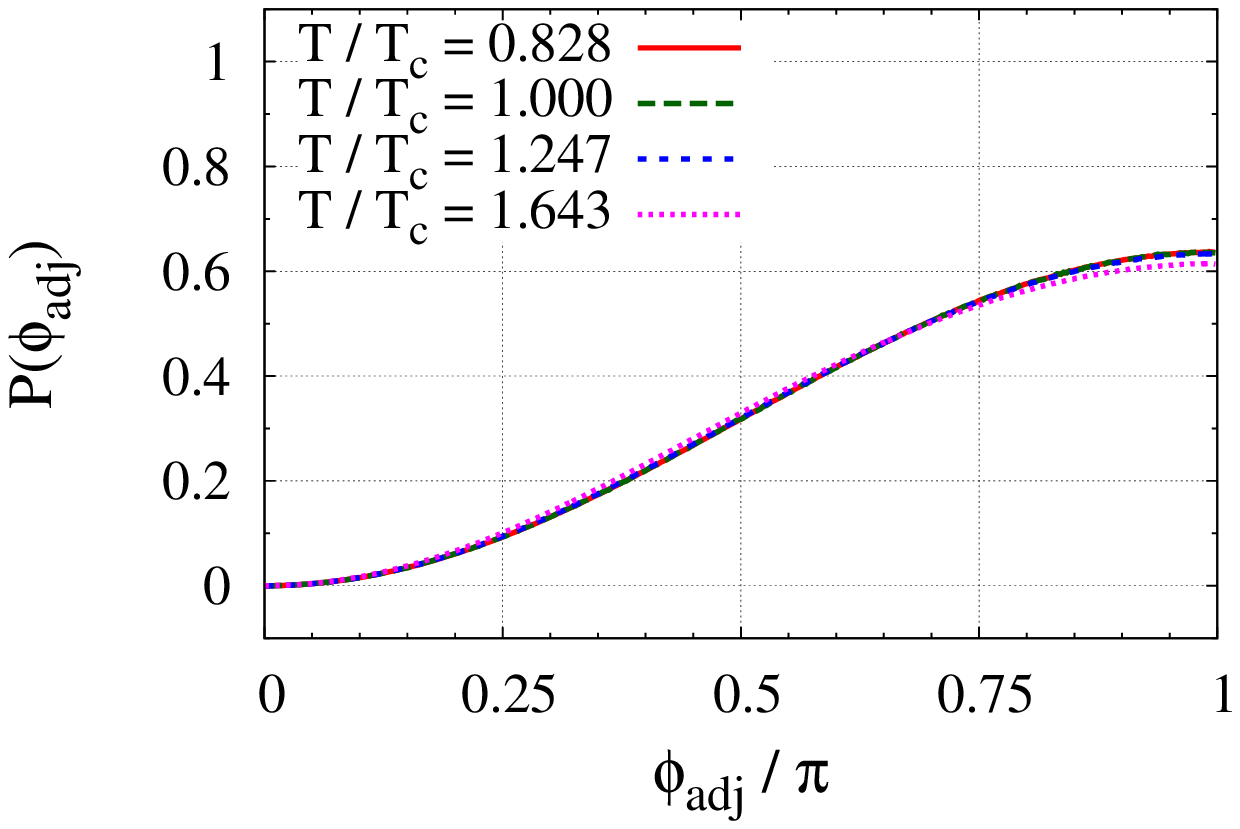}
\includegraphics*[width=\linewidth]{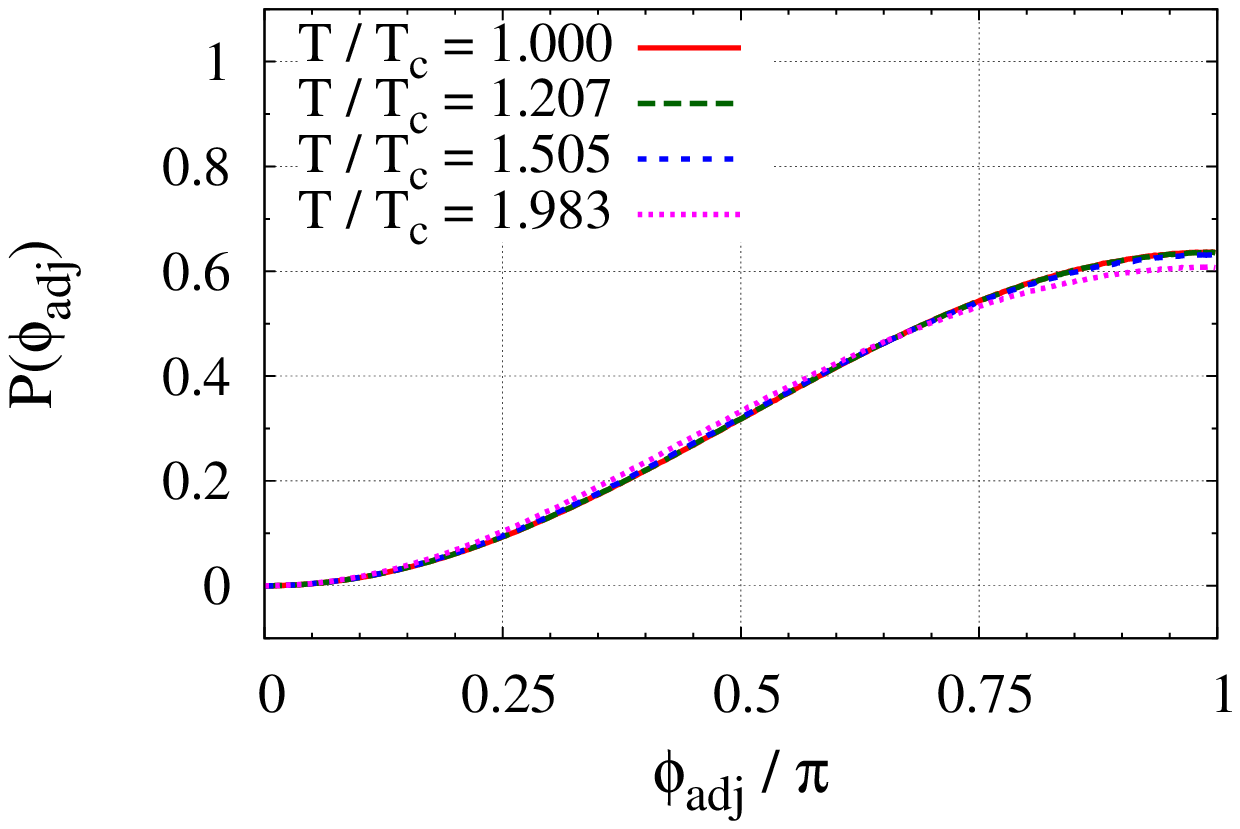}
\caption{Distribution of phase of SU(2) Wilson loop eigenvalues in
adjoint representation at $\beta=2.577856$ (top) and $\beta=2.635365$ (bottom). 
The different temperatures correspond to $N_t=12,10,8,6$ respectively.
\label{fig:su2evadj1}}
\end{figure}

\begin{figure}[h]
\includegraphics*[width=\linewidth]{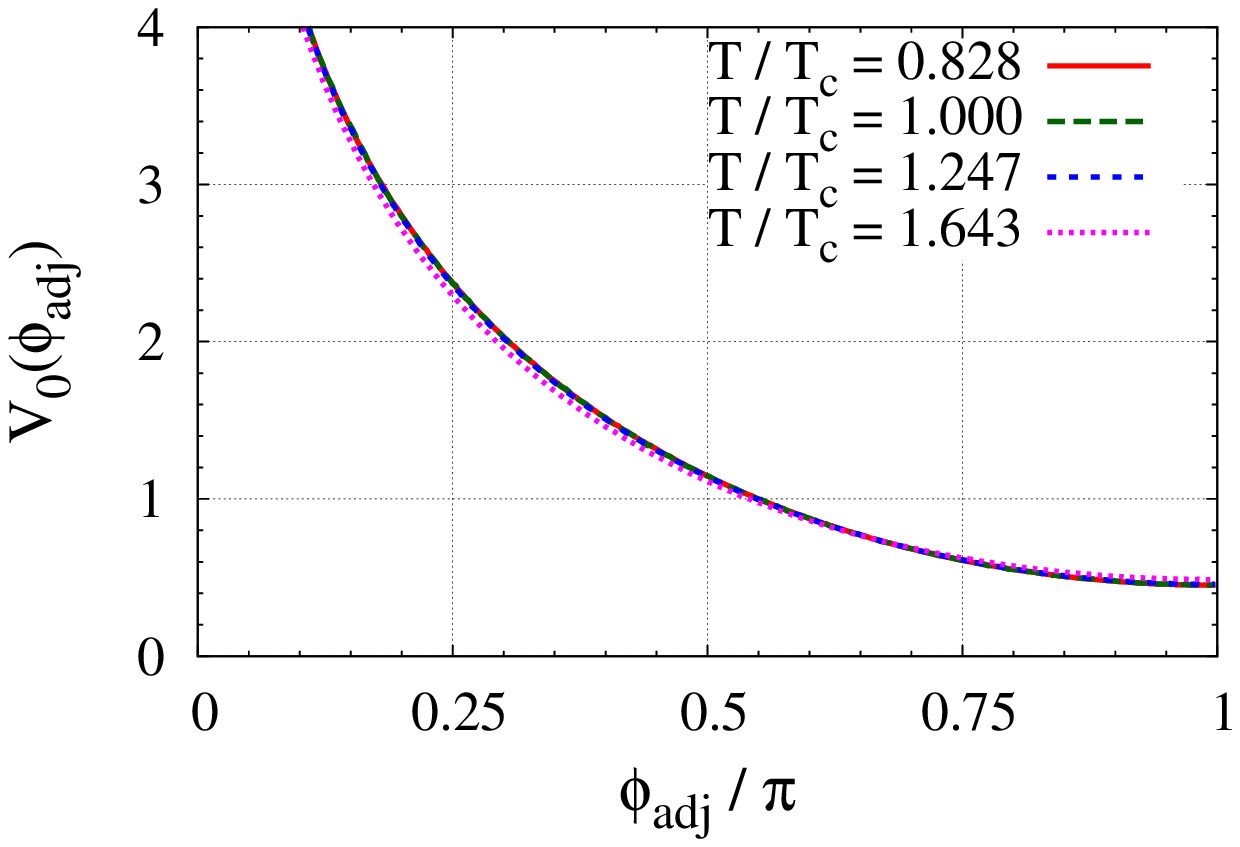}
\includegraphics*[width=\linewidth]{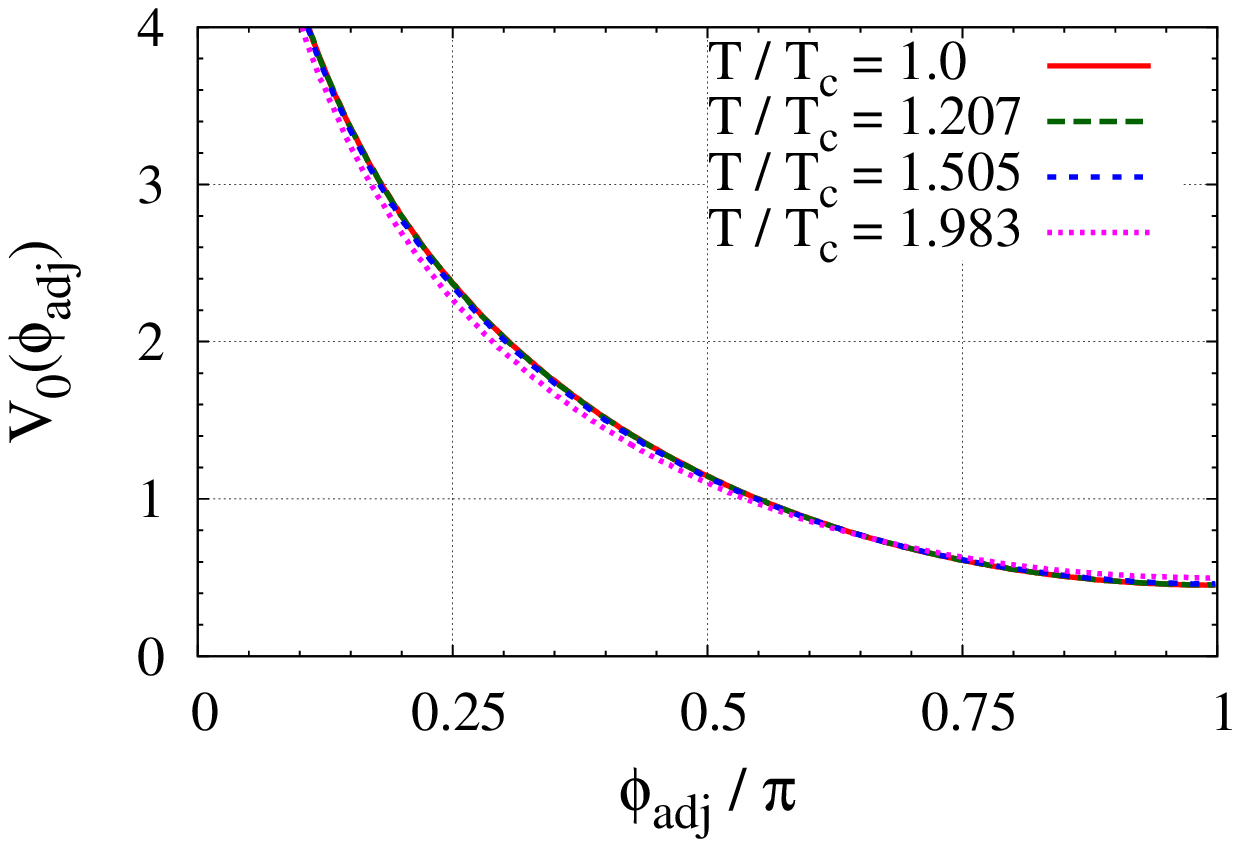}
\caption{Constraint effective potential of phase of SU(2) Wilson loop eigenvalues in
adjoint representation at $\beta=2.577856$ (top) and $\beta=2.635365$ (bottom). 
The different temperatures correspond to $N_t=12,10,8,6$ respectively.
\label{fig:su2evadj2}}
\end{figure}

From the same ensembles, we can also construct 
distributions $P(\tr\boldl^a)$ (here without
any normalization factors before the trace) for the
adjoint loop as well as for its eigenvalues.
The results are shown in Figs.~\ref{fig:su2adj1} and
\ref{fig:su2evadj1}, respectively. Figs.~\ref{fig:su2adj2} and
\ref{fig:su2evadj2} show the corresponding potentials $V_0=-\ln P$.
What is striking is that temperature effects appear
to be small. Nevertheless, a slight suppression of $P(\phi)$
around $\phi=\pi$ can be observed with increasing temperature.

We parametrize these results in a way similar to those for the
fundamental representation. We begin by transforming the
parametrization of $P(\phi)$ given in Eq.~(\ref{eq:Vand6}) to the
adjoint representation. Once this is achieved, $P(\tr\boldl)$ can be
easily obtained.  To approach this problem we require a one-to-one
mapping $\phi^f \mapsto \phi^a$ from the fundamental to the adjoint
phase. To construct such a mapping a few subtleties arise.
Eq.~(\ref{eq:funadjoint}) implies that
\be
\cos^2(\phi^f) = \frac{1}{2}\left(1+\cos(\phi^a)\right)~.\label{eq:funadjointphase}
\ee
Completing one full rotation in the fundamental phase $\phi^f$ (from
$\phi=-\pi$ to $\phi=+\pi$) generates two full cycles of the square
cosine on the left-hand side of Eq.~(\ref{eq:funadjointphase}),
whereas one rotation of the adjoint phase $\phi^a$ is only one cycle
of the right-hand side.  Both sides are symmetric around $\phi=0$,
thus it is sufficient to consider only $\phi^f \in [0,\pi]$
which can be mapped to one full rotation of the adjoint phase.
Second, the left-hand side possesses an inflection point at
$\phi^f=\frac{\pi}{2}$. Thus the mapping must be constructed
piece-wise in the intervals $\phi \in [0,\frac{\pi}{2}]$ and
$\phi \in [\frac{\pi}{2},0]$. Taking positive roots in
Eq.~(\ref{eq:funadjointphase}) and considering that
\begin{displaymath}
   +\sqrt{\frac{1}{2}\left(1+\cos(x)\right)} = \left\{
     \begin{array}{ll}
       \cos\left(\frac{x}{2}\right)  & : x \in [0,\pi]\\
       -\cos\left(\frac{x}{2}\right) & : x \in [\pi,2\pi]
     \end{array}
   \right.
\end{displaymath} 
we obtain
\begin{displaymath}
   \cos(\phi^f) = \left\{
     \begin{array}{ll}
       \cos\left(\frac{\phi^a}{2}\right)  & : \phi^f \in [0,\ldots,\pi/2]\\
       -\cos\left(\frac{\phi^a}{2}\right) & : \phi^f \in [\pi/2,\ldots,\pi]
     \end{array}
   \right.
\end{displaymath} 
and thus
\begin{displaymath}
   \phi^f = \left\{
     \begin{array}{ll}
       \frac{\phi^a}{2} & : \phi^f \in [0,\ldots,\pi/2]\\
       \pi - \frac{\phi^a}{2} & : \phi^f \in [\pi/2,\ldots,\pi]
     \end{array}
   \right.
\end{displaymath} 
This is a one-to-one mapping, covering the entire
range of both $\tr\boldl^f$ and $\tr\boldl^a$ and which
bijectively maps $\tr\boldl^f \mapsto \tr\boldl^a$ .

\begin{figure}[h]
\includegraphics*[width=\linewidth]{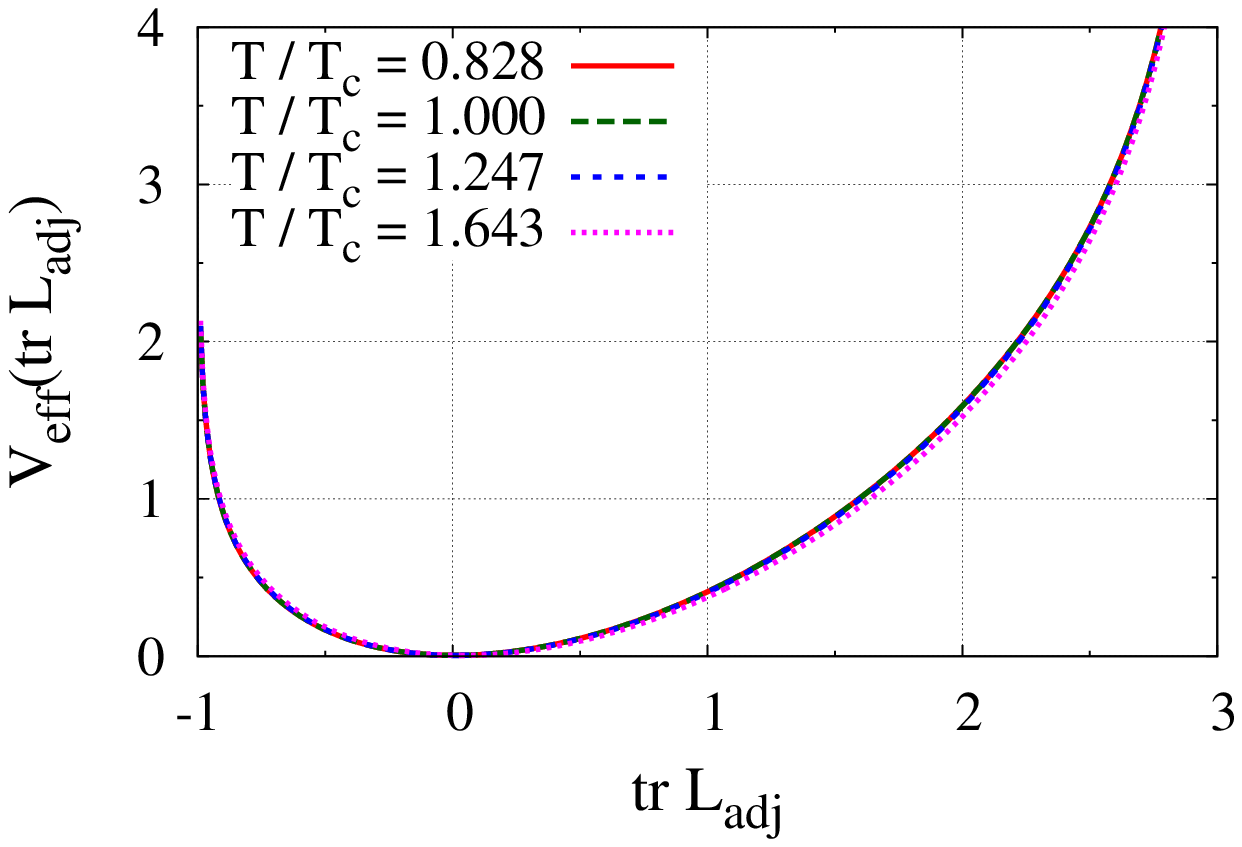}
\includegraphics*[width=\linewidth]{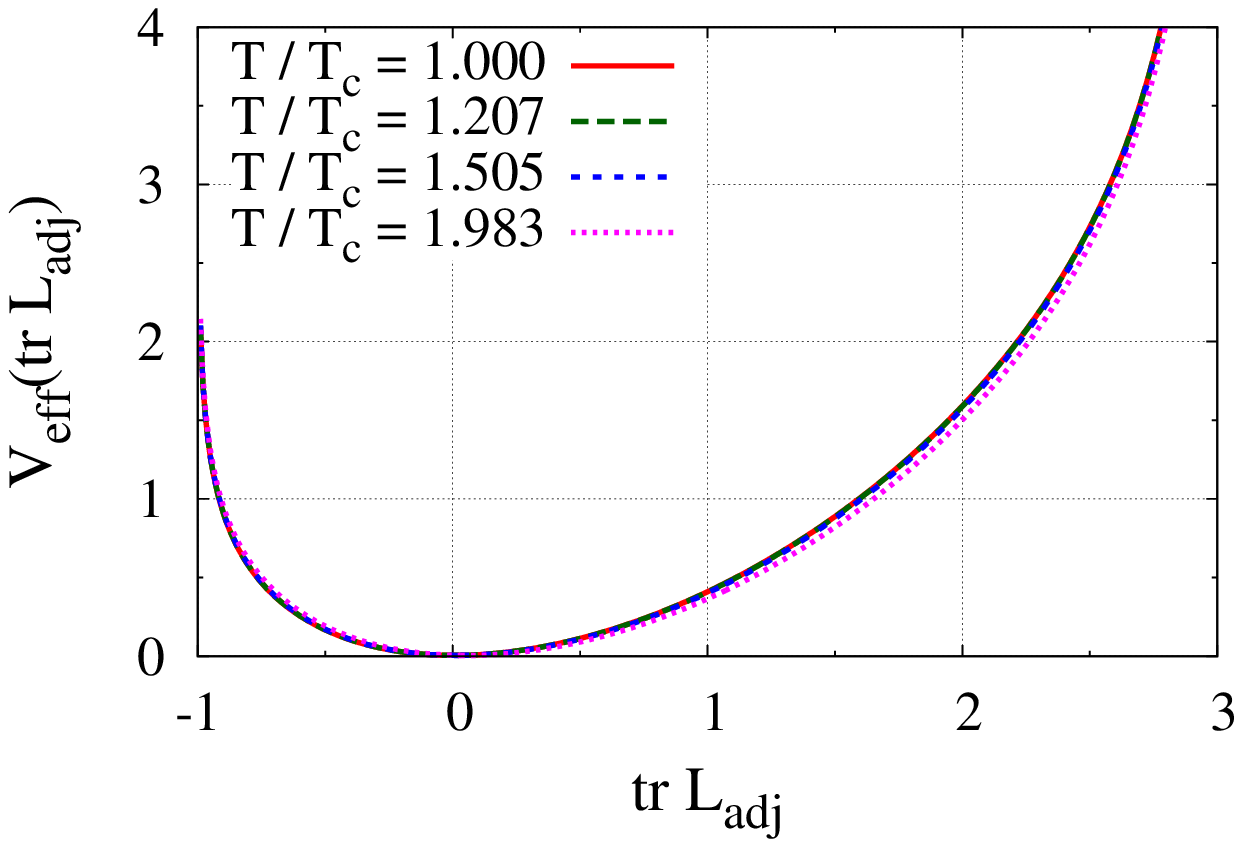}
\caption{Effective potential of SU(2) Polyakov loop in
adjoint representation at $\beta=2.577856$ (top) and $\beta=2.635365$ (bottom). 
The different temperatures correspond to $N_t=12,10,8,6$ respectively.
\label{fig:su2adj2eff}}
\end{figure}

\begin{figure}[h]
\includegraphics*[width=\linewidth]{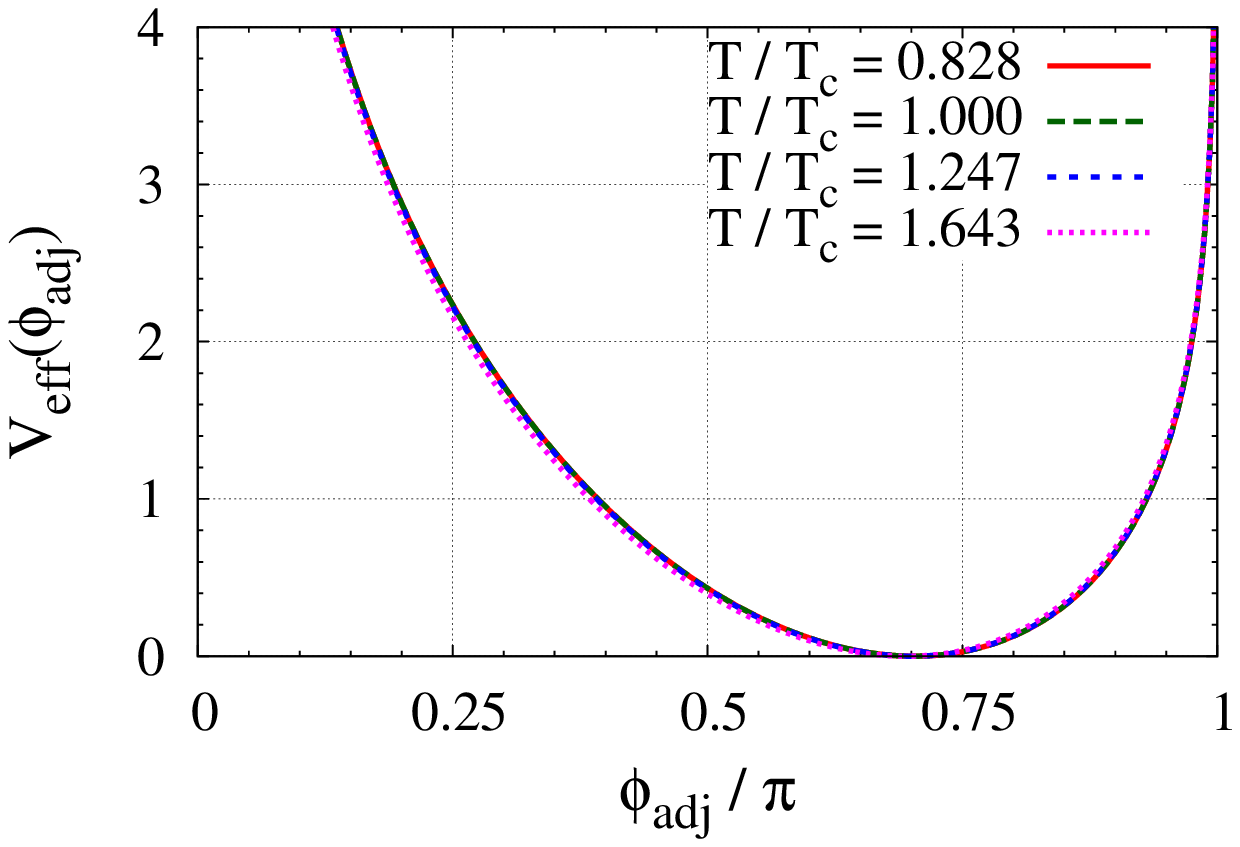}
\includegraphics*[width=\linewidth]{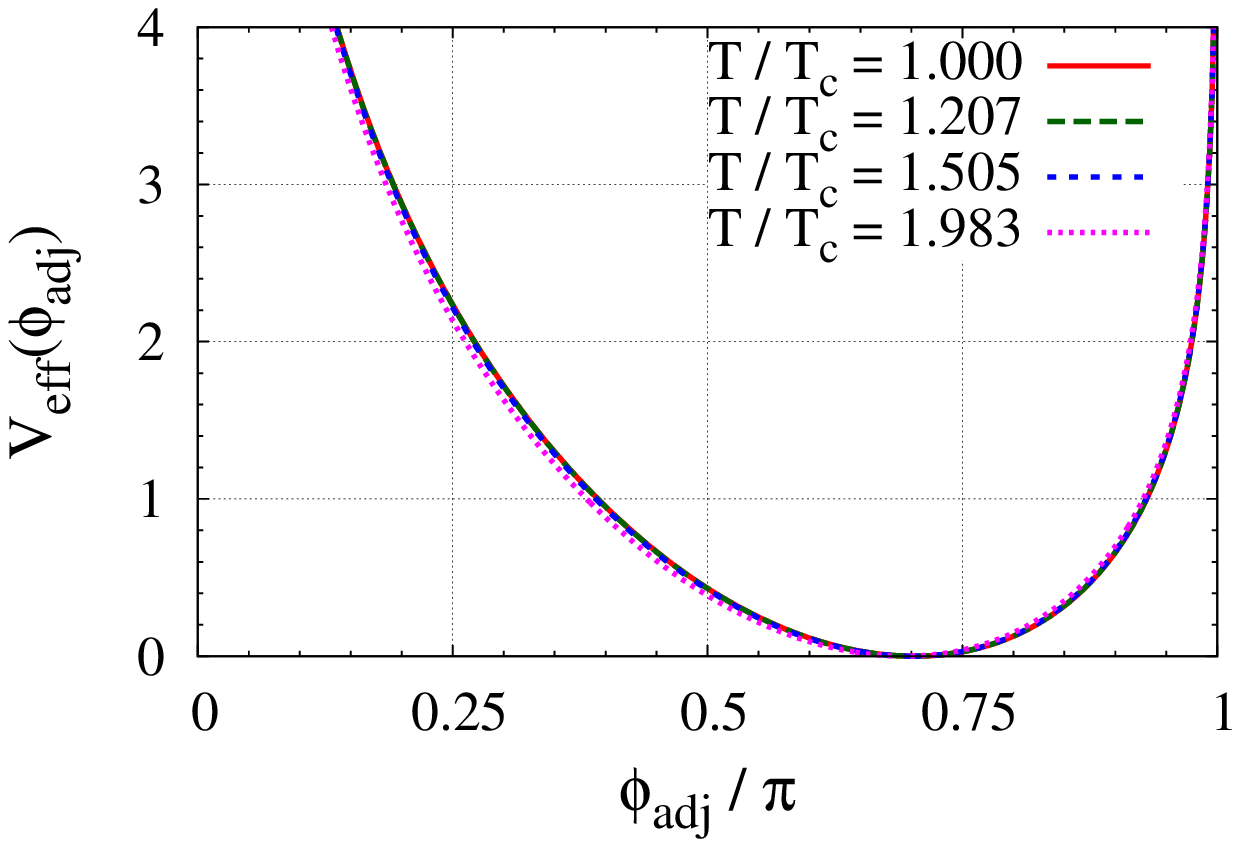}
\caption{Effective potential of phase of SU(2) Wilson loop eigenvalues in
adjoint representation at $\beta=2.577856$ (top) and $\beta=2.635365$ (bottom). 
The different temperatures correspond to $N_t=12,10,8,6$ respectively.
\label{fig:su2evadj2eff}}
\end{figure}

We can now carry out the transformation according to
\bea
\int_{0}^{\pi}d\phi^f\,P(\phi^f)&=&\int_{0}^{\frac{\pi}{2}}d\phi^f\,P(\phi^f)
+\int_{\frac{\pi}{2}}^{\pi}d\phi^f\,P(\phi^f)\notag\\
&=&\int_{0}^{\pi} P(\phi^a)d\phi^a~.
\eea
We obtain for the adjoint phase $\phi^a$
\bea
P(\phi^a)&=&P^{(T_c)}(\phi^a) \exp\left(-a(T)-c(T)\cos^2\left(\frac{\phi^a}{2}\right)\right) \notag\\
&&\times \cosh\left(b(T)\cos\left(\frac{\phi^a}{2}\right)\right)~,\label{eq:phia_distrib1}
\eea
with
\be
P^{(T_c)}(\phi^a)=\frac{1}{\pi}\sin(\phi^a)\sqrt{\frac{1-\cos(\phi^a)}{1+\cos(\phi^a)}}=\frac{2}{\pi}\sin^2\left(\frac{\phi^a}{2}\right)~.
\ee
Likewise, for the adjoint loop $\tr\boldl^a$ we obtain
\bea
P(\tr\boldl^a)&=&P^{(T_c)}(\tr\boldl^a)\exp\left(-a(T)-\frac{c(T)}{4}(1+\tr\boldl^a)\right)\notag\\
&&\times \cosh\left( \frac{b(T)}{2} \sqrt{1+\tr\boldl^a} \right)~,
\eea
with 
\be
P^{(T_c)}(\tr\boldl^a)=\frac{1}{2\pi}\sqrt{\frac{3-\tr\boldl^a}{1+\tr\boldl^a}}~.
\ee
The simulation results (Figs.~\ref{fig:su2adj1} and
\ref{fig:su2evadj1}) are again reproduced to such accuracy that the
curves are virtually indistinguishable. We carry out the Legendre
transformations~(\ref{eq:Legendre2}) to obtain
$V_{\textrm{eff}}(\tr\boldl^a)$ and $V_{\textrm{eff}}(\phi^a)$. The
results are shown in Figs.~\ref{fig:su2adj2eff} and
\ref{fig:su2evadj2eff}. Unfortunately, approximating these results
via the saddle-point method is not possible due to the lack of a
proper expansion point.
%

\begin{figure}[h]
\includegraphics*[width=\linewidth]{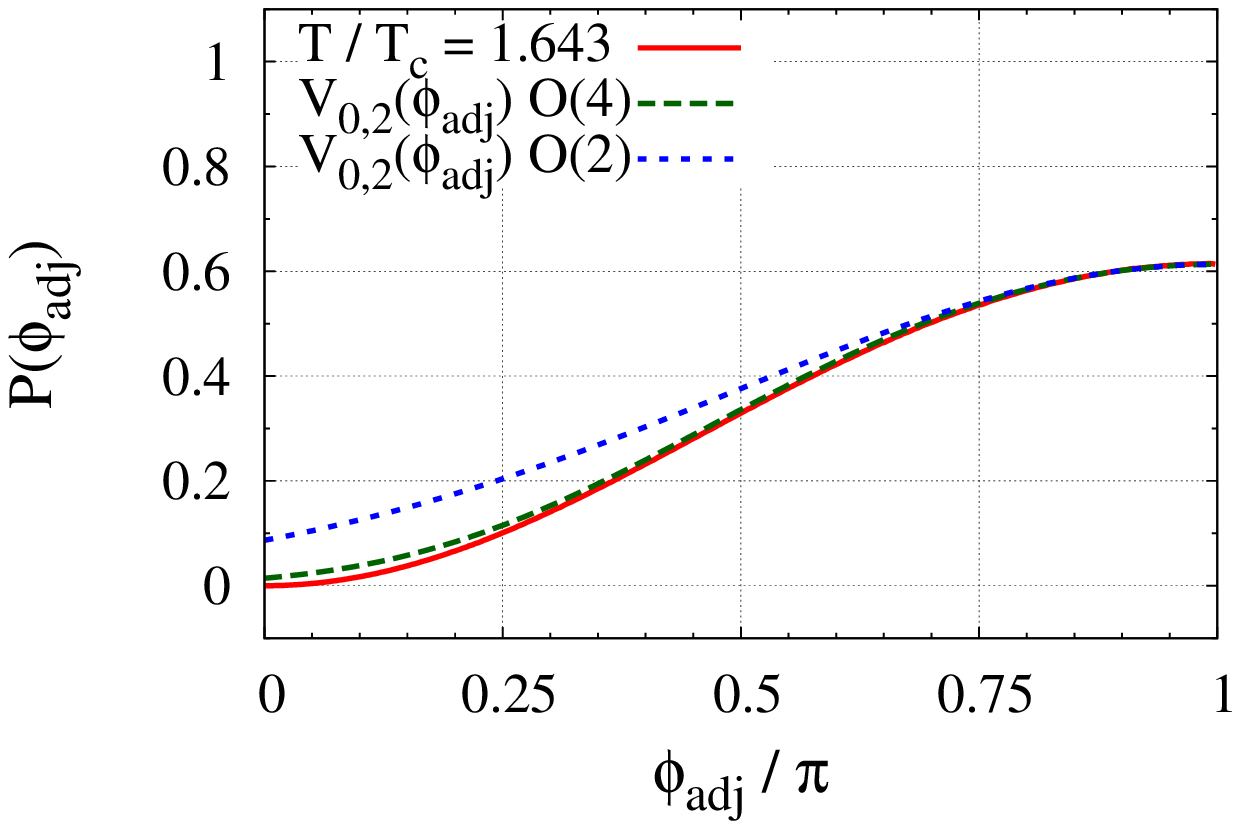}
\includegraphics*[width=\linewidth]{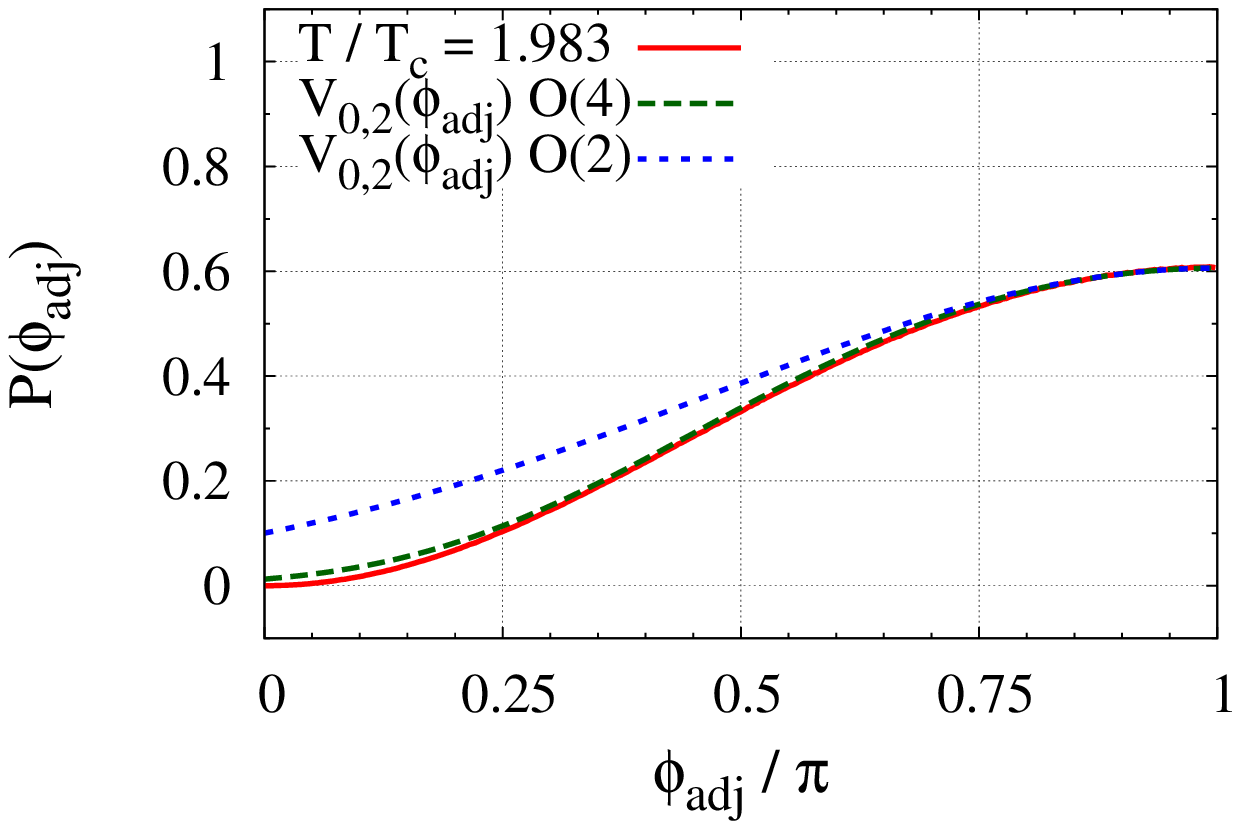}
\caption{Distribution of phase of SU(2) Wilson loop eigenvalues in
adjoint representation at $\beta=2.577856$ (top) and $\beta=2.635365$ (bottom). 
$N_t=6$ results compared to Taylor expansions.
\label{fig:su2evadj_taylor}}
\end{figure}

We again obtain a Taylor expansion of the classical potential
$V_0(\phi^a)=-\ln P(\phi^a)$.
Expanding (\ref{eq:phia_distrib1}) around $\phi^a=\pi $ to $\calo(4)$ yields
\bea
V_0(\phi^a)&\approx&-\ln\left(\frac{2}{\pi}\right)+a(T)\notag\\
&&+\left(1+c(T)-\frac{c(T)}{3}-\frac{b^2(T)}{2}\right) 
\left(\frac{\phi^a-\pi}{2}\right)^2\notag\\
&&+\left(\frac{1}{6}+\frac{b^2(T)}{6}+\frac{b^4(T)}{12}\right)
\left(\frac{\phi^a-\pi}{2}\right)^4 \label{eq:phia_distrib2}
\eea
In Fig.~\ref{fig:su2evadj_taylor} a comparison
of~(\ref{eq:phia_distrib2}) and the simulation for $N_t=6$ is shown. A
fourth order expansion reproduces the data with high accuracy. An
expansion around $\phi=0$ is not possible due to the logarithmic term
in the potential.

\section{Summary and Outlook}

In this work we simulated pure SU(2) gauge theory at finite temperature,
for two different fixed values of $\beta$ and hence of the lattice spacing.
We changed the temperature by changing the time-like extent $N_t$ 
of the lattice. For each temperature we computed the classical potential 
of the bare Polyakov loop and of the phase of the eigenvalues of the thermal
Wilson loop (wrapped around the periodic boundary of the time direction)
in the fundamental and adjoint representation, using a histogram method.
From these potentials, we obtained the effective potential via
Legendre transformation. We obtained model functions for the classical 
potentials for each case (three parameters are required) and investigated
how the parameters depend on temperature and lattice spacing. We discussed
how the models can be approximated by Taylor expansions and how they can
be converted into one another. We showed
how the effective potential of the fundamental loop can be approximated
to high precision by analytical expressions, when a next-to-leading order 
saddle expansion is used to compute the Legendre transformation. 

Several conclusions may be drawn from our work: The most obvious
conclusion is that both the classical and effective potentials of the bare
loop and its eigenvalues are well described
by simple models, reminiscent of Landau-Ginzburg theories but with a symmetry
breaking term, where the whole dependence on temperature and lattice-spacing
is absorbed into three parameters. We have thus confirmed that the ansatz used
in Refs.~\cite{Smith:2009kp,Smith:2010phd} 
to parametrize the classical potential 
in a SU(2) matrix model of Wilson lines is valid also in the full 
gauge theory, with the same qualitative behavior of the parameters.
Also, we have shown that a next-to-leading
order approximation of the effective potential of the fundamental
loop is reasonable. 

We stress that what we have computed in this work is the effective
potential for the {\it bare} loop, which vanishes
in the continuum limit.
In a given irreducible representation of the gauge
group, a renormalized loop is extracted from the bare loop by dividing
by a renormalization constant, which depends upon the representation
\cite{Gupta:2007ax}.

One of the motivations of this work was the idea that the eigenvalues
of the Wilson line might be insensitive to renormalization.  Our
results contradict this.  Instead, the eigenvalues of the bare loop are
dominated by the effects of the Vandermonde determinant.  On the lattice,
the Vandermonde determinant arises naturally at each point in spacetime.
In the action, its contribution is 
proportional to $1/a^4$, where $a$ is the lattice spacing, and so it
diverges in the continuum limit.   In retrospect, it
is natural that the lattice effective potential for a site potential
is dominated by a Vandermonde term.

For the future, it will be interesting to investigate how the
effective potential of renormalized loops can be obtained from
our results. Also, one should extract the non-perturbative
contribution to the effective potential from our data
and compare it to model calculations. Furthermore,
our investigation should be extended to higher representations
of SU(2) and also to other groups, such as SU(3) or exceptional 
groups with a trivial center, such as G(2). 
Investigating the effects of dynamical Fermions on the results
discussed here is currently under way, using standard staggered fermions.
It would be interesting to extend this investigation to
non-zero chemical potential, which is possible for two colors,
where there is no sign problem.

\section*{Acknowledgements}

We thank David Scheffler for helpful discussions, for proof-reading and for
a script to create histograms.

This work was supported by the Deutsche Forschungsgemeinschaft within
SFB 634, by the Helmholtz International Center for FAIR within the
LOEWE initiative of the State of Hesse, by the U.S.~Department of
Energy under contracts \#DE-FG02-09ER41620 and \#DE-AC02-98CH10886, by
The City University of New York through the PSC-CUNY Research Award
Program, grant 66514-00~44, and by the European Commission,
FP7-PEOPLE-2009-RG, No. 249203. 

All results presented here were obtained using
Nvidia GeForce GTX 580 graphics cards. 

\appendix
\section{Determination of physical units}
\label{sec:physunits}
To fix the physical 
units for the results discussed in this work we use
a standard method which determines the lattice spacing $a$ 
through measurements of the string tension in lattice units 
at zero temperature (which in practice means lattices of large
volume which are either hypercubic or have $N_t\gg N_x$). In this
appendix we give a short summary of the method (which
is derived in detail in several textbooks) and present
numerical results for $\beta(a)$ and related quantities. Of course the running
of the coupling with the cut-off scale has been investigated 
both analytically and numerically for both two- and three-color
gauge theory, in many previous works (see e.g. 
\cite{Bali:1992ru,Michael:1992nj,Luscher:1993gh,Necco:2001gh,Cucchieri:2002py,Bloch:2003sk} 
and references therein). 
Our main motivation for repeating
such an investigation is to make use of the advanced computing
power of the GPUs to measure at larger $N_x$ and $N_t$
and with larger statistics as was previously possible and thus
reduce systematic and statistical errors. All
physical units quoted in this work implicitly refer to the
results presented in this appendix.

In pure gauge theory the static quark-antiquark potential is 
known \cite{Luscher:2002qv} to be well described by
\be 
V(r)=A+\frac{-\pi}{12}(1/r)+\sigma r +\calo(1/ r^2).\label{eq:quarkpot3}
\ee
The linear coefficient $\sigma$ is the string tension. On the
lattice we can measure the potential in lattice units,
\emph{i.e.} $a V(a n_x)$, through the use of rectangular
Wilson loops $W(n_x,n_t)$ in the $x-t$ plane (our definition
of $W$ implies that the trace is taken). These are related to the potential by
\be 
\langle W(n_x,n_t) \rangle = e^{-n_t a V(a n_x)}
\left(1+\calo \left( e^{-n_t a \Delta E} \right) \right)~,\label{eq:quarkpot1}
\ee
where $\Delta E$ is the energy gap between the ground state and the
first excited state of a quark-antiquark pair. Since the second
term is exponentially suppressed with $n_t$, in the limit of large
$n_t$ the potential can be directly
obtained through the use of \emph{Creutz ratios}
\be 
\ln \frac{\langle W(n_x,n_t)\rangle}{\langle W(n_x,n_t+1)\rangle} \approx a V(a n_x)~.\label{eq:quarkpot2}
\ee
Since Eq. (\ref{eq:quarkpot3}) is written in lattice units as
\be 
a V(a n_x)=a A+\frac{-\pi}{12}(1/n_x)+\sigma\, a^2\, n_x + \calo(1/ (a n_x )^2) ~,\label{eq:quarkpot4}
\ee
it is clear that what we actually obtain from a $\chi^2$ fit of
the data is the dimensionless product $a^2\sigma\equiv y$. 
Fixing the string tension $\sigma$ to its physical $T=0$ value
(or some other definite value $\sigma_0$) then determines $a$
for a given $\beta$. The lattice spacing $a$ depends on the choice of
$\sigma$ of course, 
and for theories where this is not directly related to observable
physics (as in SU(2) gauge theory) 
one in fact often leaves $\sigma$ undefined,
which implies that $a$ and all derived quantities are
expressed in units of an external parameter.

The crucial point here is that the $\beta$ dependence of the lattice spacing
is in fact temperature independent. This implies that
once $a(\beta)$ is known at $T=0$, this fixes the temperature
(through $T=1/a N_t$) and the physical units of all measured
quantities for any given 
$N_t$ and $\beta$. 

To improve the signal-to-noise ratio for our string tension measurements
we employ a method known as APE smearing \cite{Albanese:1987ds}. 
This method consists of repeatedly replacing all gauge links 
$U_\mu(n)$ with averages of the form
\be 
\tilde U_\mu(n)=(1-\alpha)U_\mu(n)+\frac{\alpha}{6}
\sum_{\mu \neq \nu} C_{\mu \nu}(n)~,\label{eq:apesmear1}
\ee
where $C_{\mu \nu}(n)$ represents the staple matrix
\bea 
C_{\mu \nu}(n)&=&U_\nu(n) U_\mu(n+\hat{\nu}) U_\nu^\dagger(n+\hat{\mu} ) \nonumber \\
&+&U_\nu^\dagger(n-\hat{\nu}) U_\mu(n-\hat{\nu}) U_\nu(n-\hat{\nu}+\hat{\mu})~,\label{eq:apesmear2}
\eea
and re-unitarizing the link (projecting back to $SU(N)$) after each iteration.
The crucial effect of this procedure is to enhance the ground state overlap. APE smearing is applied to a copy of the lattice (the Monte-Carlo thus remains unaffected) on which the Wilson loops are subsequently measured. A good choice for the parameter $\alpha$ is known 
\cite{Cardoso:2010di,Geles:2011hh,Morningstar:2003gk} to be
\be
\alpha = 1 - \frac{1}{1+6w}~;~ w = 0.2 ~\rightarrow~ \alpha=0.5454 \ldots~. 
\ee
We adopt this choice as well and apply the smearing $25$ times before 
measuring Wilson loops on a given lattice configuration.

\subsection{Results}

\begin{figure}[h]
\includegraphics*[width=\linewidth]{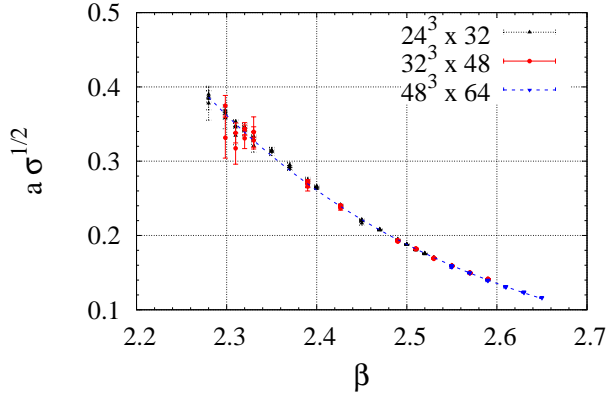}
\caption{\label{fig:runcoupl1}
The running of the lattice spacing with the gauge coupling, computed
from $n_t=9,10,11,12$ Creutz ratios on different lattices. 
The line represents $1/R(\beta)$ 
(Eqs. (\ref{eq:runcouplfit1},\ref{eq:runcouplfit2})).}
\end{figure}

We measure $a^2 \sigma$
on lattices of sizes $24^3 \times 32$, $32^3 \times 48$ and $48^3 \times 64$,
for $\beta$ values ranging from $2.28$ to $2.64$
using (\ref{eq:quarkpot2}) and (\ref{eq:quarkpot4}).
Sample sizes range from
measurements on each site of several hundreds 
(for the larger $\beta$ values) to several tens of thousands
(for the smaller $\beta$ values) of independent (smeared) gauge 
field configurations. From these results we obtain $a \sqrt{\sigma}$,
which is shown in Fig.~\ref{fig:runcoupl1}. Wilson loops are measured
up to spatial and timelike extents of $N_x/2$ and $N_t/2$ respectively.
We find consistency with the data sets quoted in Ref.~\cite{Fingberg:1992ju}.

One must consider potential sources of systematic errors. Firstly there are 
finite volume and discretization errors, for 
which we check by comparing
the results obtained from lattices of different dimensions. We find that our
different choices give consistent results in $\beta$ ranges where they
overlap (at least two different lattices overlap in any given region),
therefore we consider these errors to be under control. 
Second, one must take care to choose the timelike extent $n_t$ 
of the Wilson loops large enough, such that the contribution from
the exited states to the Creutz ratios (\ref{eq:quarkpot2}) can be neglected. 
This is tricky since the signal diminishes with rising $n_t$.
We find that we can obtain, with reasonably small statistical errors,
consistency between $n_t=9$ and $n_t=10$ on the two smaller lattices
for each value of $\beta$.
On the largest lattice consistent results are obtained on $n_t=11,12$.
Fig.~\ref{fig:runcoupl1} plots both respective choices for each lattice
as the same dataset. We consider this source of error to be under control as well. The last,
and potentially most severe, source of systematic errors are the higher
order contributions to the potential (\ref{eq:quarkpot3}), which one
must account for. We find that an excellent fit of the data
can be obtained by discarding higher order terms and restricting
the fit to $n_x \geq 2$. Alternatively one could introduce an additional
term $\sim B/r^2$, or throw away more of the short-distance results.
Although we find that the resulting string-tension measurements are much noisier
if either of the latter two options is chosen, we find evidence
that using $n_x \geq 2$ introduces a systematic error such that the string
tension is over-estimated (and thus the temperature is under-estimated)
for the region $\beta \lesssim 2.45$ if higher-order terms are excluded. 
We choose to combine two sets of
string tension data as our final results. One with $n_x \geq 2$ where the next term of
(\ref{eq:quarkpot3}) is included and one without such a term but fitting only to $n_x \geq 3$.
We find that they are consistent. We take this as evidence that the systematic
errors from higher-order terms are small for our final results. The data points shown in the figures in this section represent the measurements with the
smallest errorbars. To avoid cluttering of the figures, we refrain from showing additional points which were obtained but come with huge errorbars. 
All points were used however to determine the parameters of the model functions discussed in the following.

\begin{figure}[h]
\includegraphics*[width=\linewidth]{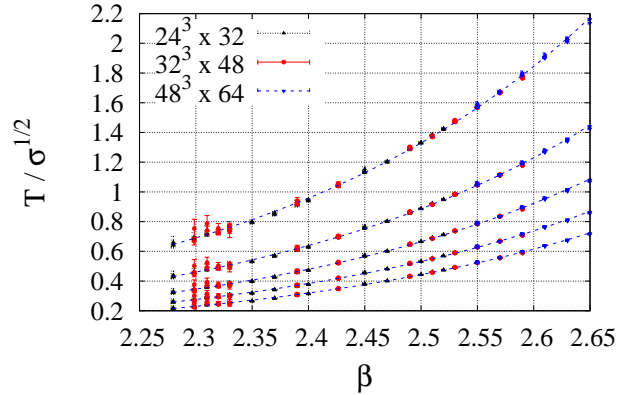}
\caption{\label{fig:runcoupl2}The physical temperature for 
$N_t=4,6,8,10,12$ (top to bottom).
The lines represent $R(\beta)/N_t$ 
(Eqs.~(\ref{eq:runcouplfit1},\ref{eq:runcouplfit2})).}
\end{figure}

Our main goal is to obtain the $\beta$ dependence of $T/T_c$ and its inverse 
for different $N_t$. This can be achieved if both $T(\beta)$ for a given $N_t$
and $\beta_c(N_t)$ are known. Furthermore,
it is desirable to obtain not only
a finite set of values but model-functions for each of these, valid
over a large as possible range. A reasonable
approach is to attempt to fit the data with known perturbative 
formulae, possibly with 
additional correction terms. 

From $a \sqrt{\sigma}$ we directly obtain 
$T/\sqrt{\sigma}=1/(a \sqrt{\sigma} N_t)$. The results are plotted
in Fig.~\ref{fig:runcoupl2} for different $N_t$. 
We find that it is possible to model our results 
by taking the inverse of the leading logarithmic term of 
the perturbative running-coupling formula (see e.g. \cite{Bloch:2003sk}).
Defining
\be
R(\beta)=\exp\left(\frac{\beta-d}{b} \right)~,\label{eq:runcouplfit2}
\ee
and using
\be
a \sqrt{\sigma}=\frac{1}{R(\beta)}~,\quad\frac{T}{\sqrt{\sigma}}N_t=R(\beta)~.\label{eq:runcouplfit1}
\ee
we find that our data shown in 
Figs.~\ref{fig:runcoupl1} and  \ref{fig:runcoupl2}
are well described for the entire range of
$\beta$ considered with the parameters 
\be
d= 1.98(1)  \quad,~b= 0.305(6) ~.\label{eq:runcouplfit3}
\ee
This is similar to what was done in Ref.~\cite{deForcrand:2001nd}
although the $\beta$ range considered here is larger. It is in fact
somewhat amazing that a leading-order formula works so well over
such a large range. It is also very convenient, 
since Eq. (\ref{eq:runcouplfit2}) can be easily inverted. 
If we restrict our fit-window to the same as was used in 
\cite{deForcrand:2001nd}, we obtain a consistent value for the parameter
$b$. If we know a $\beta_C$ for a given $N_t$ we are done,
for we can then obtain
\be
\frac{T}{T_c}=\frac{R(\beta)}{R(\beta_c)}
=\exp\left(\frac{\beta-\beta_c}{b} \right)~,\label{eq:totc}
\ee
and likewise
\be
\log\left(\frac{T}{T_c}\right)b+\beta_c= \beta~.
\label{eq:runcouplfit4}
\ee
Our numerical results for $T/\sqrt{\sigma}$ can
analogously be converted to $T/T_c$. The parameter
$\sigma$ drops out entirely. 

\begin{figure}[h]
\includegraphics*[width=\linewidth]{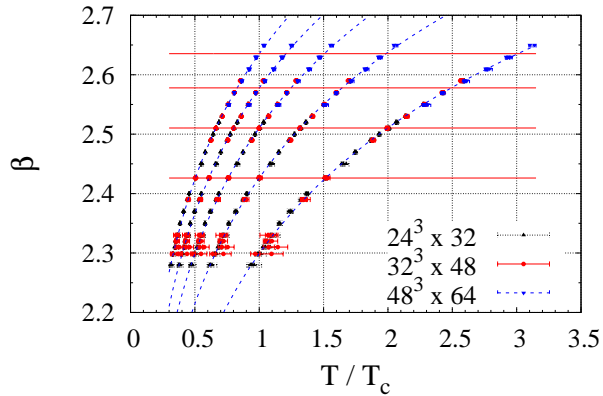}
\caption{\label{fig:runcoupl3}The running coupling for
$N_t=12,10,8,6,4$ (left to right). The horizontal
lines mark $\beta_C$ for $N_t=12,10,8,6$ (top to bottom).
The model function is defined in Eq. (\ref{eq:runcouplfit4}).}
\end{figure}

We show $\beta$ as a function of $T/T_c$ in 
Fig.~\ref{fig:runcoupl3}, both as raw data
and model-function. We collected $\beta_c$ values for
$N_t=4,5,6,8,12,16$ from 
Refs.~\cite{Fingberg:1992ju,Velytsky:2007gj,Cheng:2012rg}
and fit them to interpolate for other $N_t$ values, using 
\be
\beta_c(N_t)= a_0 + b_0 \log(N_t) - \log(\log(N_t))^{c_0}~,\label{eq:nlo_bc}
\ee
We find that the next-to-leading logarithm is necessary
to appropriately describe such a range of $N_t$. 
Our fit-parameters converge to 
\be
a_0 = 1.1579(6)~,~b_0=0.9398(1)~,~c_0=1.627(2)~,
\ee
with $\chi^2/\textrm{dofs}\approx 0.18$. 
Eq. (\ref{eq:nlo_bc}) is used in Eq. (\ref{eq:totc}) to
determine $\beta_c$ for all $N_t$ instead of the raw
data.


One last thing should be mentioned: We realize that the results
presented in this appendix go beyond what is necessary
for the study presented in the main sections. In particular,
to obtain $T/T_c$, when simulating on a lattice of time-like
extent $N_t$ at a given $\beta$ which is
known to be $\beta_c$ for a different $N_t^c$, one can simply
use the relation
\be
\frac{T}{T_c}=\frac{N_t^c}{N_t}~.\label{eq:totc_ntontc}
\ee
We conducted the investigation discussed here hoping
that it will be useful in a more general setting and
we have chosen to present $T/T_c$ values throughout this
paper as our general method provides them.
Equation (\ref{eq:totc_ntontc}) however is very useful
to us to estimate errors. It turns out that the $T/T_c$
values quoted throughout this paper differ by no more
than $\sim2\%$ from what (\ref{eq:totc_ntontc}) predicts,
which is quite satisfactory.

\end{document}